\documentclass[reprint, superscriptaddress]{revtex4-1}

\usepackage{amsmath, amssymb, amsfonts}
\usepackage{bm}
\usepackage{color}
\usepackage{float}
\usepackage{graphicx}
\usepackage{hyperref}
\usepackage{dcolumn}

\graphicspath{{images/}}

\begin{document}

\title{Embedding of biological distribution networks with differing environmental constraints}

\author{Lia Papadopoulos}
\affiliation{Department of Physics and Astronomy, University of Pennsylvania, Philadelphia, PA 19104, USA}
\author{Pablo Blinder}
\affiliation{Sagol School of Neuroscience, Tel-Aviv University, Tel Aviv 6997801, Israel}
\affiliation{Neurobiology Department, George S. Wise Faculty of Life Sciences, Tel Aviv University, Tel Aviv 6997801, Israel}
\author{Henrik Ronellenfitsch}
\affiliation{Department of Physics and Astronomy, University of Pennsylvania, Philadelphia, PA 19104, USA}
\author{Florian Klimm}
\affiliation{Mathematical Institute, University of Oxford, Oxford OX2 6GG, UK}
\affiliation{Systems Approaches to Biomedical Science Doctoral Training Centre, University of Oxford, Oxford OX1 3QU, UK}
\author{Eleni Katifori}
\affiliation{Department of Physics and Astronomy, University of Pennsylvania, Philadelphia, PA 19104, USA}
\author{David Kleinfeld}
\affiliation{Department of Physics, University of California San Diego, La Jolla, CA 92093, USA}
\author{Danielle S. Bassett}
\affiliation{Department of Bioengineering, University of Pennsylvania, Philadelphia, PA 19104, USA}
\affiliation{Department of Electrical \& Systems Engineering, University of Pennsylvania, Philadelphia, PA 19104, USA}
\email{dsb@seas.upenn.edu}

\date{\today}

\begin{abstract}
Distribution networks -- from vasculature to urban transportation systems -- are prevalent in both the natural and consumer worlds. These systems are intrinsically physical in composition and are embedded into real space, properties that lead to constraints on their topological organization. In this study, we compare and contrast two types of biological distribution networks: mycelial fungi and the vasculature system on the surface of rodent brains. Both systems are alike in that they must route resources efficiently, but they are also inherently distinct in terms of their growth mechanisms, and in that fungi are not attached to a larger organism and must often function in unregulated and varied environments. We begin by uncovering a common organizational principle -- Rentian scaling -- that manifests as hierarchical network layout in both physical and topological space. Simulated models of distribution networks optimized for transport in the presence of fluctuations are also shown to exhibit this feature in their embedding, with similar scaling exponents. However, we also find clear differences in how the fungi and vasculature balance tradeoffs in material cost, efficiency, and robustness. While the vasculature appear well optimized for low cost, but relatively high efficiency, the fungi tend to form more expensive but in turn more robust networks. These differences may be driven by the distinct functions that each system must perform, and the different habitats in which they reside. As a whole, this work demonstrates that distribution networks contain a set of common, emergent design features, as well as tailored optimizations.
\end{abstract}

\maketitle

\section{Introduction}

Many complex networks \cite{Newman:2010aa} that describe physical and biological systems are \textit{spatial} in the sense that the nodes and edges are embedded into real space \cite{Barthelemy:2011}. A classic example is a transportation network, where stations correspond to nodes and where physical routes, such as roads or railways, correspond to network edges \cite{Kurant:2006a, Sperry:2016a, Latora:2002a}. Some biological systems can also be thought of as transportation networks, for example transmitting electrical signals as is the case in collections of neurons or large-scale brain areas \cite{Bullmore:2009aa}. The physical nature and spatial embedding of these systems often imposes costs on network wiring, making spatially long-distance connections improbable, and thereby constraining the network's topology \cite{Gastner:2006b}. Pressures that compete against wiring minimization include those driving network efficiency and robustness. Tradeoffs between these desirable network features can differ across systems, and directly inform the design of optimal spatial distribution networks \cite{Gastner:2006a, Newman:2006a, Bohn:2007a, Durand:2007a, Banavar:2000a, Li:2010a, Banavar:1999aa, Hu:2013a}.

Biological distribution systems often grow and evolve in particularly interesting environments, and are indeed subject to the competing pressures of maintaining low material costs while achieving high efficiency and robustness. Distribution systems can constitute an entire organism -- such as mycelial fungi -- in which the physical cords making up the organism can be represented as edges in a network, and in which branching, fusion, or end points among those cords can be represented as nodes in a network \cite{Heaton:2012a, Fricker:2007a, Lee:2016a}. These systems appear to strike an intermediate balance between cost and efficiency, enabling the organism to achieve competing goals. In addition, their network architecture changes and adapts over time in a way that tends to strengthen beneficial qualities \cite{Bebber:2007a,Fricker:2007a,Fricker:2008b,Fricker:2008c, Fricker:2009a, Boddy:2009a}. Alternatively, distribution systems can form only a small part of an organism -- as is the case with cortical vasculature -- in which a pial network on the surface of the brain routes blood to penetrating arterioles, that in turn supply the underlying tissue \cite{Shih:2015a, Blinder:2013aa}. The pial surface network from rodent brains can also be represented as a network, in which edges are vessels, and nodes represent either branching points among those vessels or a branch to a penetrating arteriole \cite{Blinder:2010a}. This system forms a robust backbone of loops that can withstand damage and re-rout flow in the presence of occlusions \cite{Blinder:2010a, Schaffer:2006a}.

In both the fungal and vasculature systems, two-dimensional, planar distribution networks must transport fluid and nutrients efficiently in the face of constraints on the total amount of material that they can support. But in spite of these commonalities, the two networks exist and have evolved in inherently different environments, which may directly impact the sorts of evolutionary pressures that the network might experience. For example, the main role of the vasculature network is to transport blood to and from tissue that is part of a larger organism. On the other hand, for fungi, the network is the organism itself and is not necessarily constrained to serve or occupy a set region of space. Moreover, brain vasculature resides in a controlled environment within the confines of the skull, whereas fungal networks live in and must adapt to often unprotected and varied environmental conditions \cite{Heaton:2012a,Boddy:1999b, Boddy:2007a,Rotheray:2008a}. In addition, while directed flow and growth are known to be important in both vasculature and fungi, the mechanisms of long-distance transport of nutrients and maturation are different in the two systems \cite{Heaton:2012a}. However, while the different roles, habitats and function of vasculature \emph{vs.} mycelial systems should directly affect their network architecture, little is known about what these similarities or differences in structure might be.

To better understand the network configurations of different types of distribution systems, we utilize a set of methodologies to investigate both the \textit{physical} and \textit{topological} structure of mycelial fungi and rodent brain vasculature networks. We first uncover architectural commonalities in the network organization of both systems in the form of previously unidentified physical and topological scaling relationships that indicate hierarchical structure, and constrained spatial embedding. We also show that a theoretical model for distribution networks -- based on an optimization of transport efficiency in the face of fluctuations in load -- is able to reproduce the same design features, and thus provides a mechanism through which the empirically observed spatial and topological network organization might arise. Yet despite these shared features of network structure, the two types of networks also display differences in the complexity of their architectures, suggesting that each system may be differently optimized for cost, efficiency, and robustness. By comparison to a set of spatially-informed null models, we explicitly examine the \textit{relative} relationships between these three quantities, and compare and contrast the associated tradeoffs across organisms. This analysis uncovers clear differences in how the two types of distribution networks balance competing goals. Taken together, our work uncovers several common organizational principles across distribution systems, but also identifies critical distinctions that we hypothesize may be markers for the network's function and reflect the environment in which that function must be performed.
\begin{figure}
	\centering
	\includegraphics[width = \linewidth]{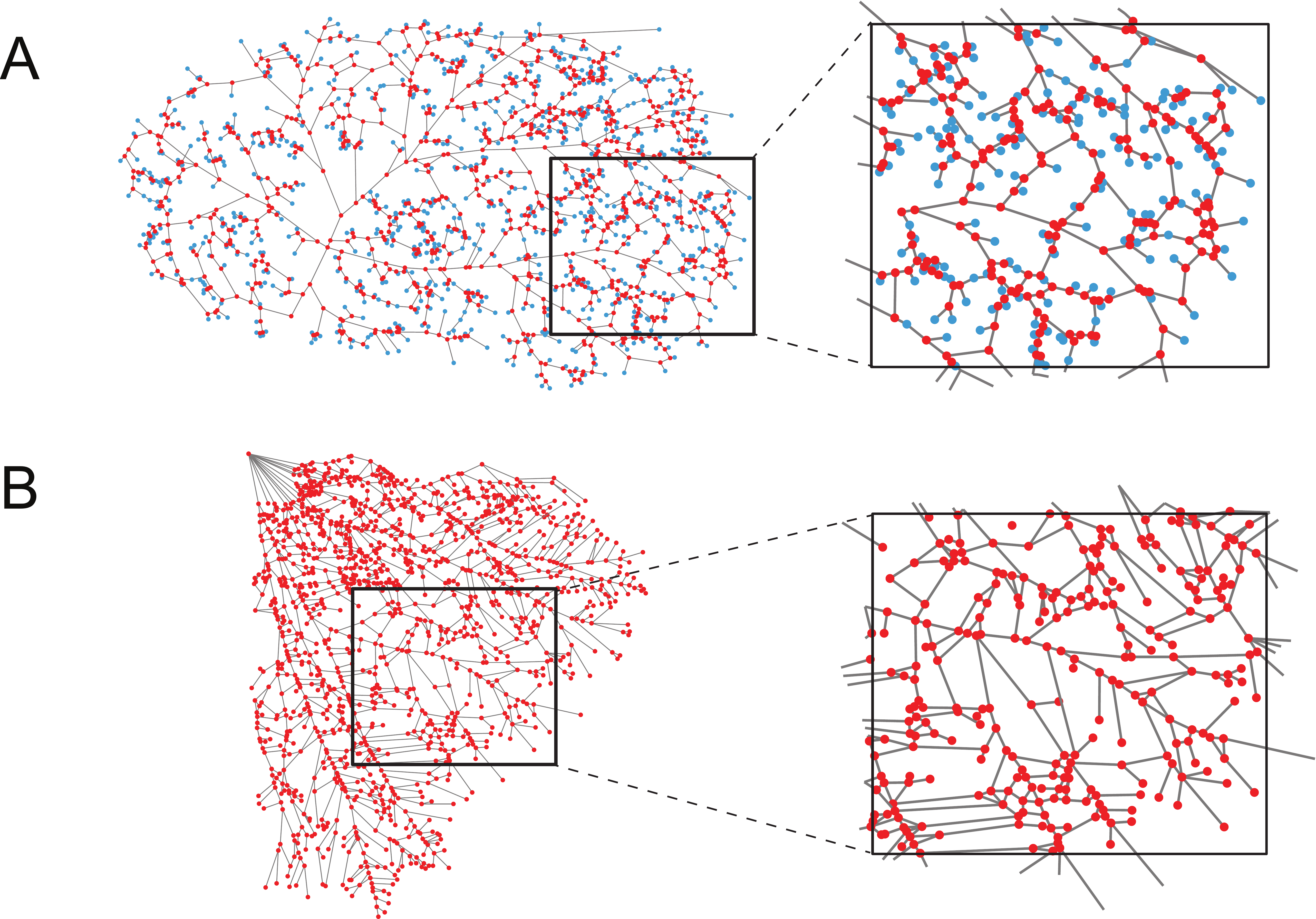}
	\caption{\textbf{Distribution networks from rodent vasculature and mycelial fungi.} \emph{(A)} An example of the vasculature network on the surface of a rat brain in the region of the middle cerebral artery. Vessel segments are represented as edges and connect branching points on the surface backbone (red nodes) or penetrating arterioles (blue nodes). \emph{(B)} An example of the network of the fungus \textit{Phallus impudicus}. The cords making up the organism are represented as edges and connect to form branching, fusion, or end points.}
\label{f:networks}
\end{figure}

\section{Results}
\label{s:results}

\subsection{Network representation}

We study the vasculature and fungi by considering both as \textit{complex networks}, each represented by an adjacency matrix $\mathbf{A}$. In an unweighted network, $A_{ij} = 1$ if there is an edge between nodes $i$ and $j$, and zero otherwise. This results in a binary connectivity matrix that captures the topological structure of the system. In spatial networks, there is often additional information due to the embedding of the network in physical space. Here, the networks are embedded into two-dimensional space, so a given node $i$ has a spatial coordinate, $\{x_{i}, y_{i}\}$, which allows the Euclidean distance between nodes $i$ and $j$, $D_{ij}$, to be computed.

We examine the surface vasculature from the neocortex of four rats and five mice \cite{Blinder:2010a}, and three different species of mycelial fungi, including \textit{Phallus impudicus (P.I.)}, \textit{Phanerochaete velutina (P.V. 1)} (grown from five inocula), \textit{Phanerochaete velutina (P.V. 2)} (grown from a single inocula), and \textit{Resinicium bicolor (R.B.)} \cite{fungalData}. An example of the entire vasculature of the middle cerebral artery from a rat brain is shown in Fig.~\ref{f:networks}A and an example mycelial network from \textit{P.I.} is shown in Fig.~\ref{f:networks}B. More information about the data and network construction can be found in \textit{Methods} and the \textit{S.I.}.

\subsection{Distribution networks exhibit physical Rentian scaling}
\label{s:pRent}
\begin{figure}
\includegraphics[width = \linewidth]{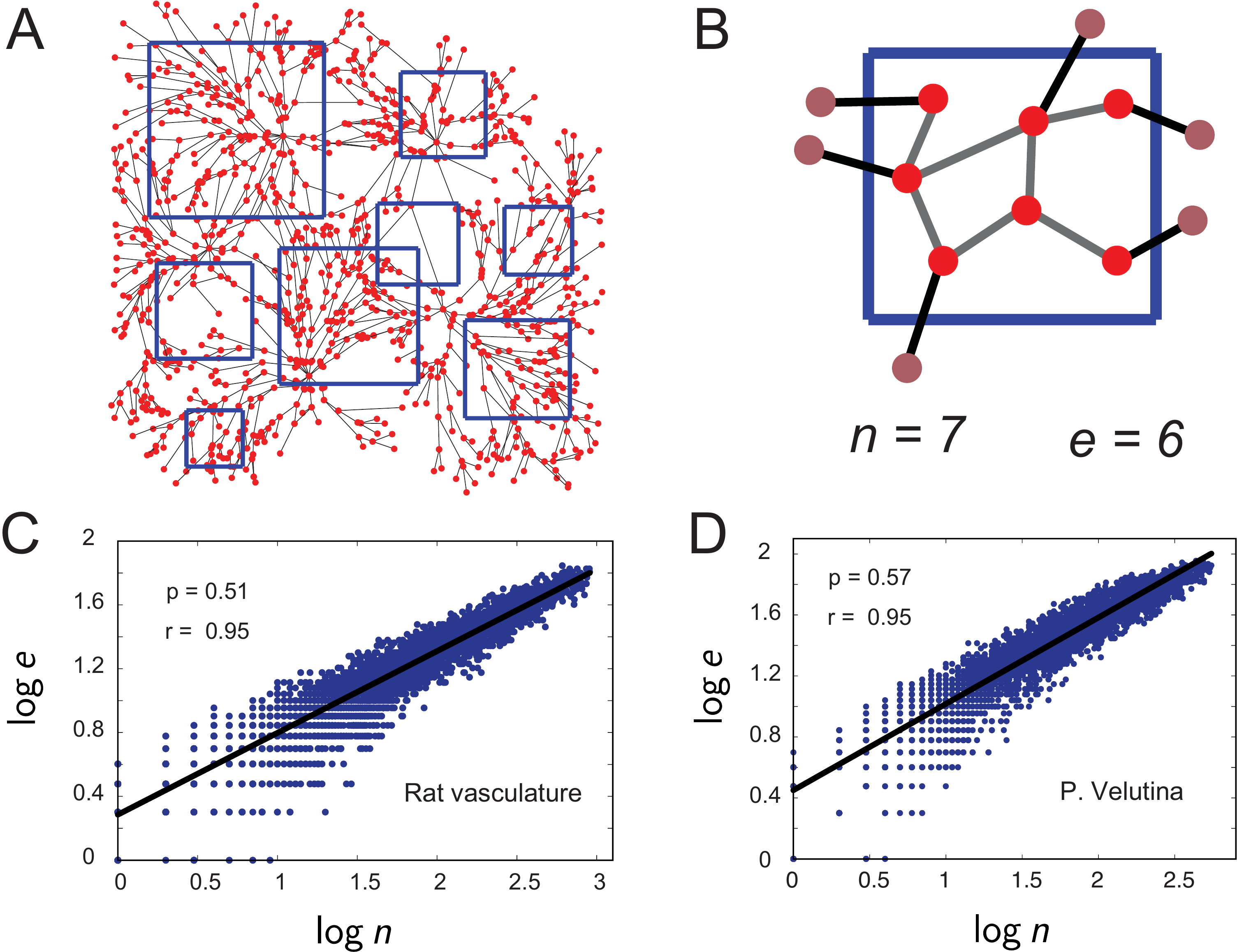}
\caption{\textbf{Physical Rentian scaling.} \emph{(A)} We test for hierarchical structure in real space by placing boxes of different length scales at random over the network, in this case, the fungus \textit{P.V. 1}. \emph{(B)} For each box, we count the number of nodes inside $n$ and the number of edges crossing the boundary $e$, and consider how these quantities scale in $\log-\log$ space. If scaling exists, the physical Rent exponent $p$ is determined from the slope of the best fit line. \emph{(C)} An example of the physical Rentian scaling relationship in the arterial network of a rat brain, and \emph{(D)} in a network of the fungus \textit{P.V 1}. In both instances, there is a strong linear relationship (high Pearson correlation $r-value$). Data for the rest of the networks are reported in the \textit{S.I.}. The lines correspond to a best fit line of the data points from which we estimate the displayed scaling exponent $p$.}
\label{f:pRent}
\end{figure}
An important question regarding biological transport systems is whether distribution networks of different types or from different organisms exhibit universal characteristics that allow for high functionality, and in particular, whether they achieve low-cost embeddings into physical space while maintaining topologies that support efficiency and robustness. To address these questions, we draw on fundamental insights from computer science developed in the 1960s in the context of very large scale integrated (VLSI) circuits \cite{Chen:2000a}. At that time, it was noted that a salient feature of computer chips -- whose abstract topologies enabled complex information flow and computational functions -- was \textit{Rentian scaling}, and that this feature allowed for a low-cost embedding of the circuitry into physical space, facilitated by generally short wires \cite{Christie:2000a, Ozaktas:1992a, Verplaetse:2001a, Greenfield:2010a,Stroobandt:2011a,Stroobandt:1998a}. Rentian scaling manifests in a power law relationship between the number of logic gates (or nodes) $G$ in a region of the network and the number of terminals (or edges) $T$ passing through that region, such that $T \propto G^{p}$, where $0 \leq p \leq 1$ is the Rent exponent. This principle of physical embedding has been observed in both artificial and natural intelligent systems \cite{Bassett:2010a}, and in general, signifies hierarchical network structure. Importantly, this scaling notion extends ideas of self-similarity (such as the fractal dimension \cite{Song:2005aa}), which have otherwise been agnostic to spatial constraints.

To determine the presence or absence of such a principle in biological distribution networks, we place square boxes of different sizes over each network (see \textit{S.I.} for details), and define $n$ to be the number of nodes within a box, and $e$ to be the number of edges crossing the boundary of a box (Fig~\ref{f:pRent}A,B). Here, $e$ is a measure of the amount of flow that can pass into or out of a given region of the network, relative to the number of nodes, or distribution sites, within that region. After laying 5000 boxes over the network, we test for the scaling relationship $e \propto n^{p}$ where $p$ is the physical Rent exponent. We observe that all networks considered show evidence of physical Rentian scaling with $p$ ranging from $\approx 0.51$ to $\approx 0.59$ (for examples of this relationship in vasculature and a mycelial system, see Fig.~\ref{f:pRent}C,D, respectively). This finding implies that the spatial layout of the distribution networks is hierarchically organized, and suggests an efficient embedding of the network into physical space in terms of short node-to-node distances. While scaling was weakest in the mouse vasculature, for all networks, we found Pearson's correlation coefficients of $r > 0.9$; $p$-value $< 0.05$ (\textit{S.I.}). To give context to the exponent values, we also report $p$ for several idealized network topologies in Table~\ref{t:rent_ideal}, including a triangular lattice, a square lattice, a hexagonal lattice, and a minimum spanning tree (averaged over several realizations). More details on the construction of each of these networks can be found in the \textit{S.I.} The vasculature networks have exponents greater than that of a minimum spanning tree or square lattice, and fall roughly in the same range of $p$ as a hexagonal lattice. Many of the fungal networks also have exponent values similar to either a hexagonal or triangular lattice, but some are outside this range and it thus appears that they cannot be explained by the physical architecture of the regular lattices.

\begin{table}[h]
\centering
\begin{tabular}{c c c c}
\hline\hline
Triangular lattice 	& 	Square lattice 		&	Hexagonal lattice	&	Spanning tree 	\\
\hline
p = 0.537			&	p = 0.500			& 	p = 0.519			&	p = 0.502 	      	 \\
t = 0.475			&	t = 0.437			&	t = 0.402			&	--- 			\\
\hline\hline
\end{tabular}
\caption{Physical and topological Rentian scaling exponents for idealized networks.}
\label{t:rent_ideal}
\end{table}

\subsection{Distribution networks exhibit topological Rentian scaling and low topological complexity}
\label{s:tRent}
\begin{figure}
\includegraphics[width = \linewidth]{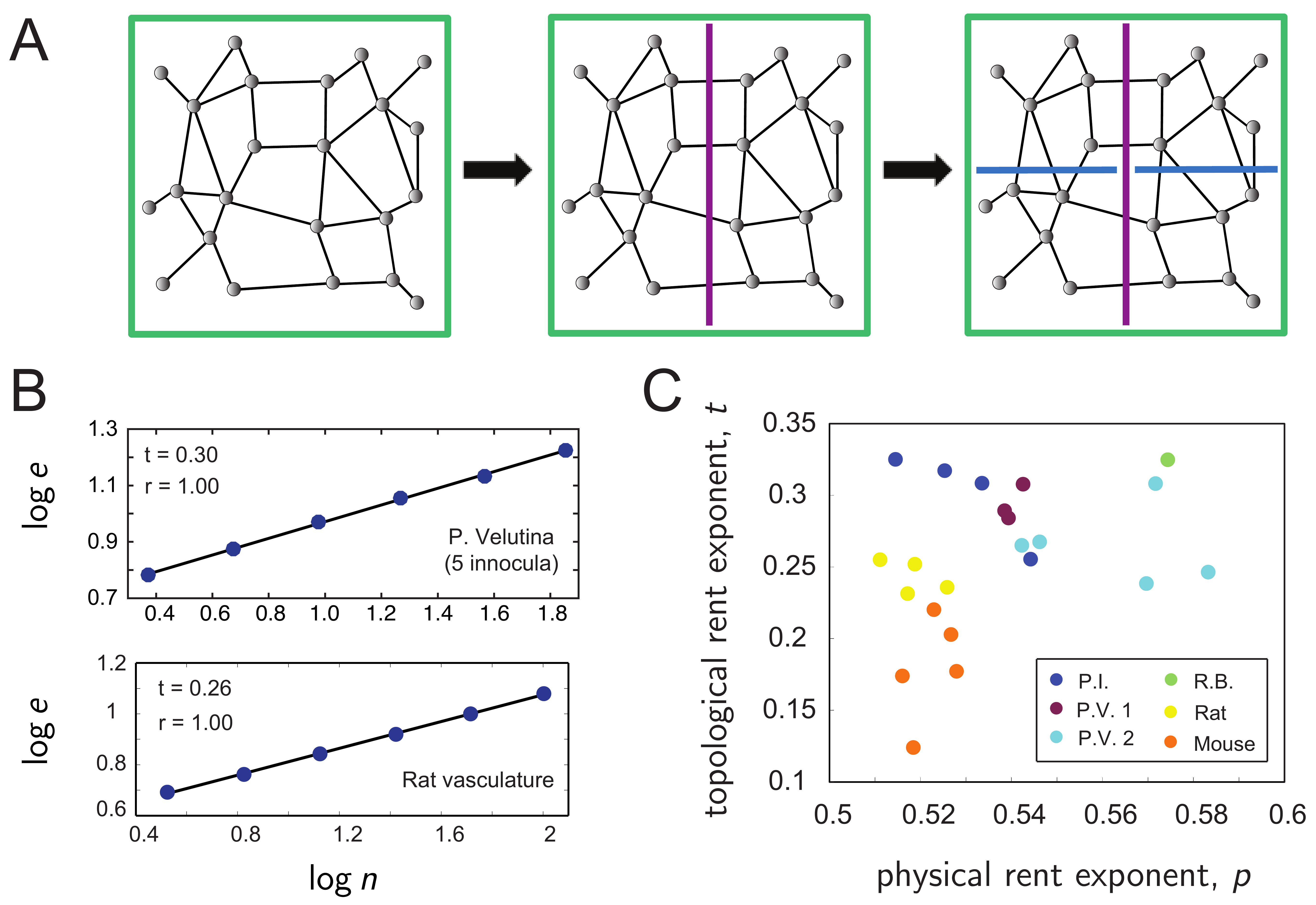}
\caption{\textbf{Topological Rentian scaling.} \emph{(A)} A schematic of the topological partitioning process on a toy network. Beginning with a single box that covers the entire network, we recursively partition the network into halves, quarters, etc., such that the number of nodes $n$ within each partition is roughly equal, and the number of edges $e$ that cross partitions is minimum. If topological scaling exists, $n$ and $e$ scale with one another in $\log-\log$ space and the topological Rent exponent $t$ is given by the slope of the best-fit line. \emph{(B)} An example of the topological Rentian scaling relationship in the vasculature from a rat brain (top), and in the fungal network \textit{P.V. 1} (bottom). For both networks, there is a strong linear relationship (high $r-value$). Data for the rest of the networks are reported in the \textit{S.I.}. The lines correspond to a best fit line of the data points from which we estimate the displayed scaling exponent $t$. \emph{(C)} For each class of networks, we compute $p$ and $t$ and plot them against one another to characterize the complexity of network structures in physical and topological space.}
\label{f:tRent}
\end{figure}
The relative ease with which a network can be embedded into physical space depends significantly on the complexity of the network's topology: low-dimensional topologies are easier to embed than higher-dimensional topologies. To probe the complexity of these distribution networks, we carried out a Rentian scaling analysis in topological space by testing for a relationship of the form, $e \propto n^{t}$, where $n$ is the number of nodes inside a topological partition of the network, $e$ is the number of edges crossing the partition boundary, and $0 \leq t \leq 1$ is the topological exponent. In a recursive min-cut bi-partitioning algorithm \cite{Karypis:1999a, hMetis}, the network is split into halves containing roughly equal numbers of nodes and configured such that the number of edges between the two topological partitions is minimum. Each half is then split again according to the same rule, and this procedure continues recursively (Fig.~\ref{f:tRent}A). We observe that all networks considered also show evidence of Rentian scaling in this abstract topological space, with $t$ ranging from $\approx 0.1$ to $\approx 0.35$ (for examples from vasculature and fungi, see Fig.~\ref{f:tRent}B). This finding further highlights the presence of self-similarity in the distribution systems, but this time, in the non-spatial architectural layout of the network. Scaling was again weakest in the mouse vasculature, but for all networks, we found Pearson's correlation coefficients of $r > 0.98$; $p$-value $< 0.05$ (\textit{S.I.}). We again compare these exponent values to those from the idealized networks in Table~\ref{t:rent_ideal}. We first note that by nature, the pure tree does not exhibit topological scaling since a single edge cut always splits the network into disconnected components; in this way, the presence of loops is required to have topological scaling. Interestingly, however, we observe that the values of $t$ for all of the lattice networks are greater than the values from all the real networks, both fungal and vasculature. Thus, the degree of topological complexity in the real distribution systems is not accurately captured (it is overestimated) by idealized lattices.


How are physical and topological scaling related to network complexity? First, the topological dimension of a network $d_{\text{T}}$ can be related to $t$ via $t \geq 1 - \frac{1}{d_{\text{T}}}$; higher topological exponents indicate higher dimensional network topology \cite{Ozaktas:1992a,Stroobandt:2011a}. Moreover, the theoretical minimum value for $p$ is related to the Euclidean dimension of the network $d_{E}$ and topological Rent exponent $t$ via $p_{min} = max(1 - \frac{1}{d_{\text{E}}}, t)$; for these 2-dimensional systems, $p_{min} = 0.5$ \cite{Verplaetse:2001a}. The theory of VLSI circuit design predicts that networks with higher values of $p$ (i.e., values further from the theoretical minimum) tend to have more complex spatial organization and wiring designs, and are often not efficiently embedded in terms of short edge lengths. Notably, here we find without exception that the topological exponent is smaller than the physical exponent, $t < p$, and that $t < 1 - \frac{1}{d_{\text{E}}} = p_{min} = 0.5$ (Fig.~\ref{f:tRent}C). These relationships imply that for the class of transport networks studied here, the embedding dimension $d_{\text{E}}$ is greater than the topological dimension $d_{\text{T}}$. This is in contrast to a much different situation in natural and artificial neural networks, for example, where the topological dimension is greater than the embedding dimension \cite{Bassett:2010a}. The fact that $d_{\text{T}} < d_{\text{E}}$ in the distribution networks is indicative of significant constraints on wiring in these systems. Though the real networks do contain redundancies and shortcuts in the form of loops, unlike idealized lattices, they also have many tree-like regions with branches ending in single nodes. This yields space-filling networks with relatively low topological complexity and material cost.


\subsection{Rentian scaling in model transport networks}

An interesting question lies in understanding what sort of construction rules and theoretical models for transport networks could give rise to the physical and topological scaling principles observed in the real data. To address this line of inquiry, in addition to characterizing the spatial and topological layout of the empirical networks, we also examine Rentian scaling in an ensemble of model networks built from a biologically inspired optimization principle. In particular, we study a set of distribution networks based on a resistor network formalism, which has been investigated previously in several works \cite{Banavar:2000a, Bohn:2007a, Durand:2007a, Katifori:2010a}. We note that the goal of this analysis is not to capture the specific details and differences in each family of network, but rather to shed light on a common mechanism that is able to capture some of the key physical and topological properties that are present in \textit{both} systems.

For the resistor network simulation, one chooses a set of nodes to be sources or sinks, and pairs of nodes are joined together by conductances. Voltage differences across nodes then drive current flow through the system, and the goal is to optimize the conductance through each edge (under a fixed total conductance cost), such that the power dissipated by the network is minimized. In order to make the model more realistic for biological distribution networks, we follow \cite{Katifori:2010a}, and optimize the networks for transport efficiency while subjecting the system to fluctuations in load (modeled as a moving sink in the network). Importantly, introducing this type of heterogeneity can produce loops in the optimal networks, a feature that appears salient in the real systems studied here, and that likely affects the observed values of the scaling exponents.

\begin{figure}
\includegraphics[width = \linewidth]{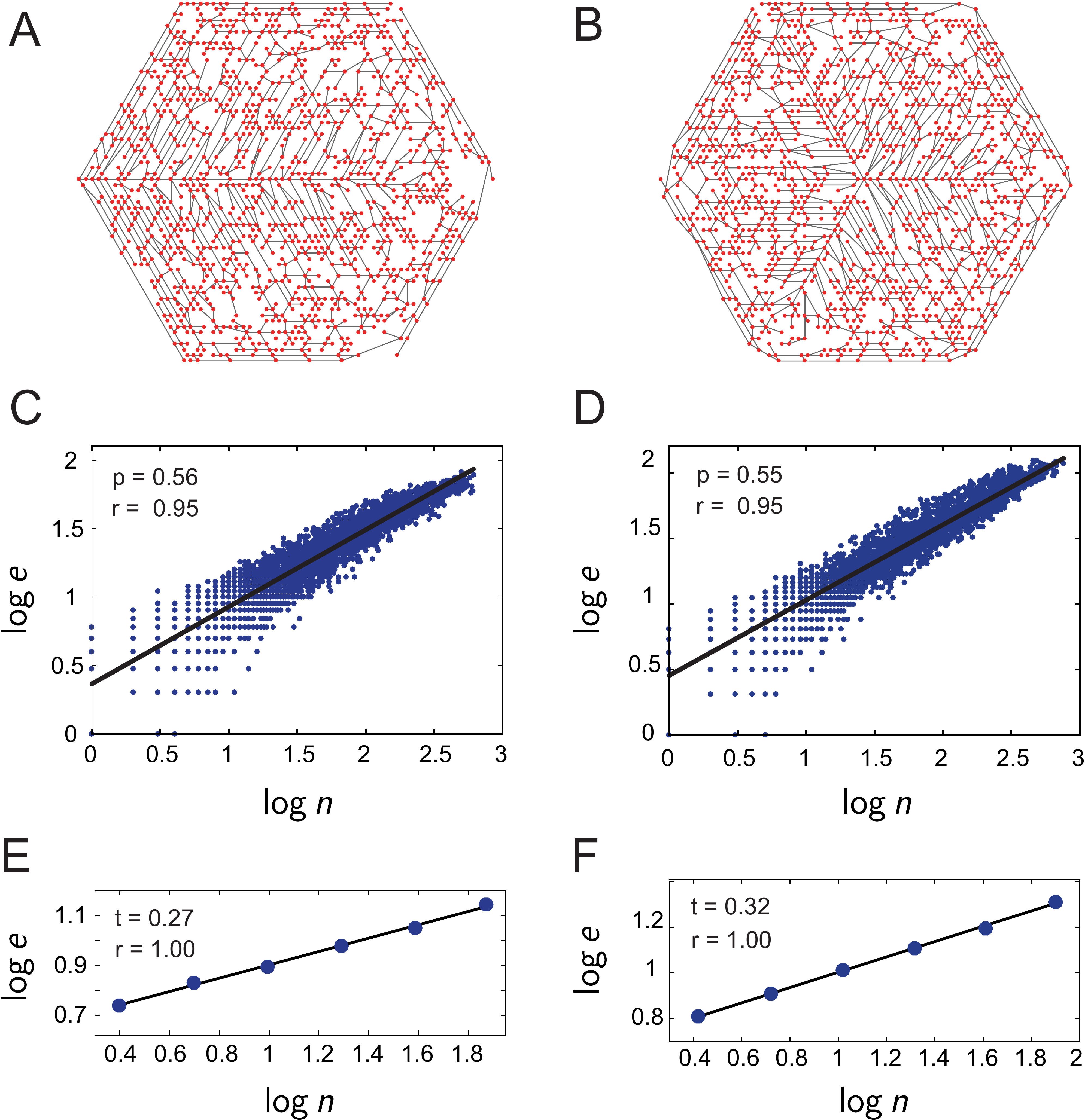}
\caption{\textbf{Rentian scaling in model networks.} Examples of model distribution networks constructed by optimizing transport in resistor-like networks subject to fluctuations in load. \emph{(A)} To simulate the rodent vasculature systems, the source was placed at the left edge. \emph{(B)} To simulate the fungi, the source was placed at the center. \emph{(C, D)} Both types of model networks exhibit Rentian scaling in physical space, with exponents in the same range as those found in the empirical data. \emph{(E, F)} Each network additionally displays strong topological Rentian scaling, with exponents close to the empirically observed values as well.}
\label{f:model_networks}
\end{figure}

To capture the organization of the vasculature and fungi, we construct two versions of the optimal transport networks that differ in terms of the location of the source node. As a model for the pial network, which branches outward from the Circle of Willis, we set the source to be at one edge of the system (without loss of generality, we pick the left edge). On the other hand, to more accurately model the fungal systems, which in the experiments grow outward from a single seed (or in the case of \emph{P.V. 1}, a set of 5 symmetrically placed seeds), the second set of simulations is run with the single source node placed in the center (Fig.~\ref{f:model_networks}A--B). After performing the simulations for network optimization, all networks were binarized by keeping only the edges with a conductance greater than a threshold value. Further details on the model formulation and network construction can be found in the \emph{S.I.}

\label{s:tRent}

After generating the model networks, we asked whether or not they showed Rentian scaling, and furthermore, if the values of the exponents were qualitatively similar to those observed in the real systems. Using the analysis described in Sec.~\ref{s:pRent} and \ref{s:tRent}, we find that the optimal transport networks do exhibit both the physical and topological form of hierarchical structure (Fig.~\ref{f:model_networks}C--F). As before, we quantify the goodness of the linear relationship between $\log{n}$ and $\log{e}$ using the Pearson correlation coefficient, $r$. Averaged over all networks, we find $r$ = 0.949 $\pm$ 5.027e-4 for the physical scaling, and $r$ = 0.998 $\pm$ 2.273e-4 for the topological scaling. Perhaps of greater interest is whether the model gives rise to scaling exponents in the same range as those estimated from the natural systems. To test this, we compute the mean physical and topological scaling exponents, $\overline{p}$ and $\overline{t}$, respectively, averaged over all simulated networks. The average values are $\overline{p}$ = 0.547 $\pm$ 0.002 and $\overline{t}$ = 0.263 $\pm$ 0.004. Comparing these quantities to the data points in Fig.~\ref{f:tRent}\emph{C}, we find that they are indeed in the correct range, lying approximately in the center of the real data. These results suggest that the resistor network formalism, in combination with the optimization of transport under fluctuations, provides a mechanism that is able to produce model networks with architectural properties similar to the empirical data.

\subsection{Tradeoffs between wiring and efficiency}
\begin{figure}
\includegraphics[width = \linewidth]{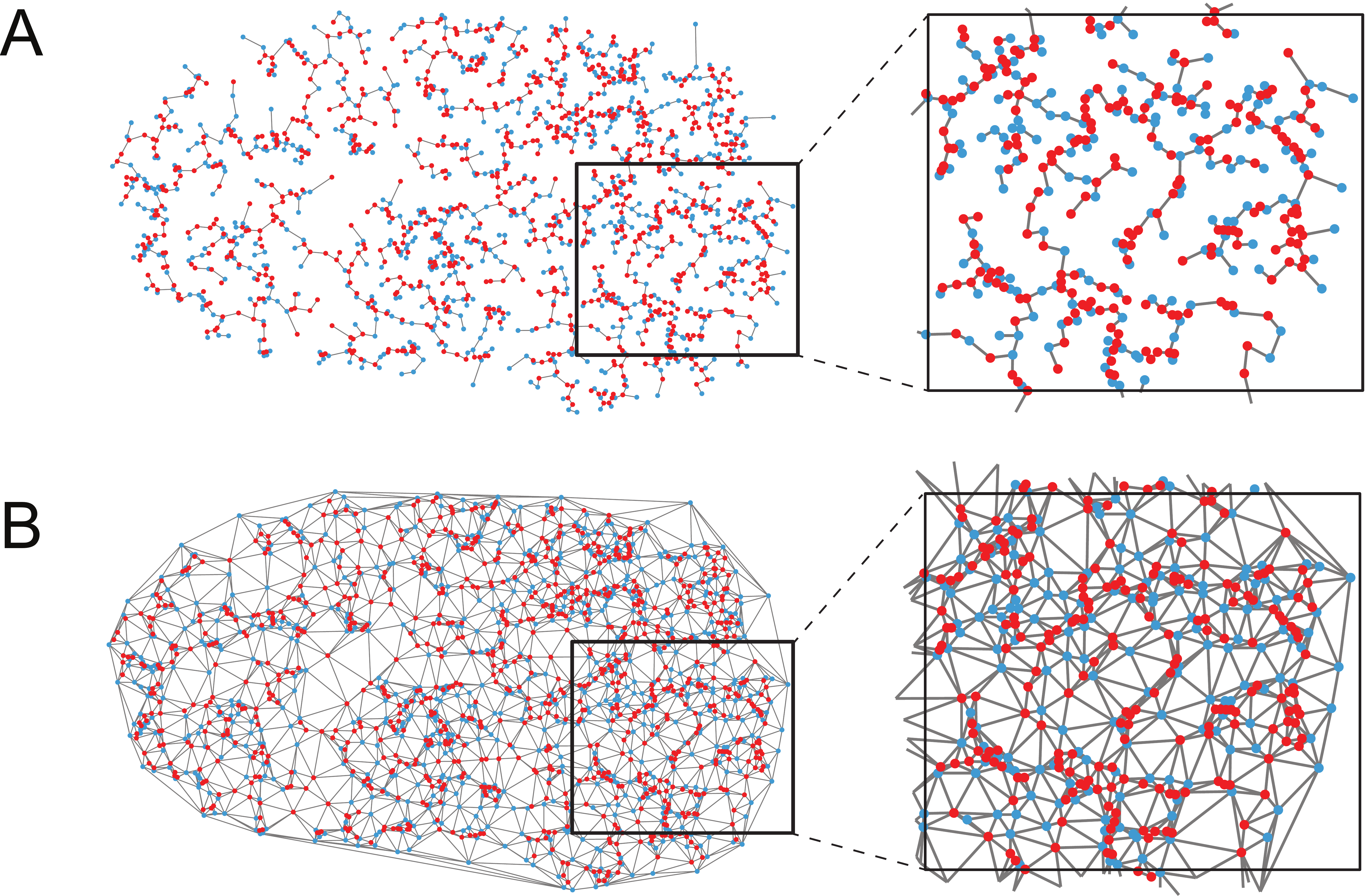}
\caption{\textbf{Construction of spatial null model networks.} \emph{(A)} The minimum spanning tree (\textit{MST}) and \emph{(B)} the greedy triangulation (\textit{GT}) for the vasculature network in Fig.~\ref{f:networks}.}
\label{f:null_models}
\end{figure}
\begin{figure*}
\includegraphics[width = \textwidth]{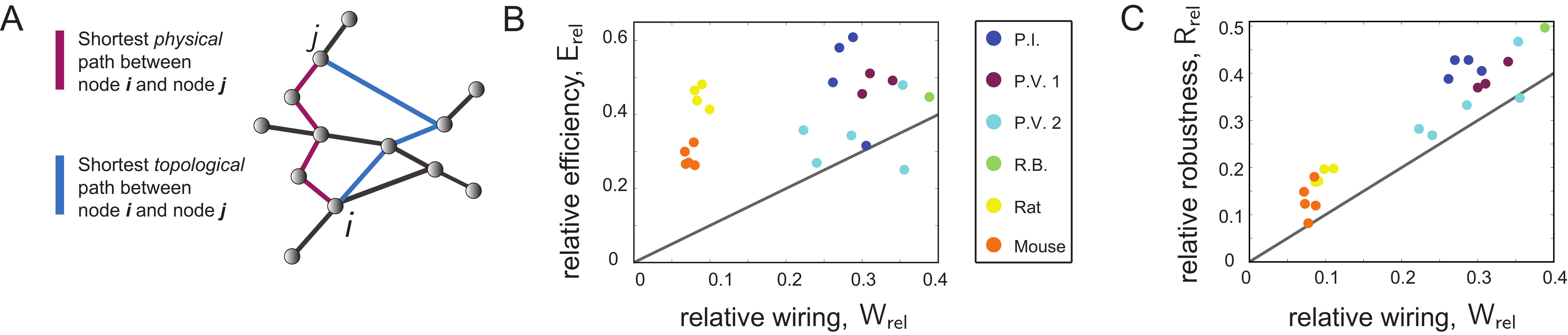}
\caption{\textbf{Vasculature and fungal networks exhibit different optimizations.} \emph{(A)} A toy network that depicts the difference between the shortest \textit{physical} path between two nodes and the shortest \textit{topological} path between the same nodes. We use the former of the two measures to obtain a spatial notion of network efficiency. \emph{(B)} A scatter plot of the relative wiring \emph{vs.} the relative efficiency for each network. The gray line corresponds to $E_{\text{rel}} = W_{\text{rel}}$. \emph{(C)} A scatter plot of the relative wiring \emph{vs.} the relative robustness for each network. The gray line corresponds to $R_{\text{rel}} = W_{\text{rel}}$. For both plots the relative quantities are determined from a comparison of each network to its own null models, allowing for a further comparison of the different types of distribution networks (vasculature and fungi) to one another.}
\label{f:wiring_eff_paths}
\end{figure*}

The results thus far paint a picture of both vasculature and fungal networks as having self-similar architecture over different spatial and topological scales, and as being characterized by low topological complexity and a physical embedding that minimizes long-distance connections. Despite these similarities, the environments and functions of these two systems suggest that they may be optimized for different structural features. To probe such differences, we explicitly examine a set of measures that are important considerations in spatially embedded, biological systems, specifically. In particular, the geometric global efficiency -- defined as the reciprocal of the harmonic mean of the shortest physical path lengths between nodes \cite{Latora:2001a} -- is a measure that reflects the routing capabilities of a network (see Fig.~\ref{f:wiring_eff_paths}A; \textit{Methods}, \textit{S.I.}). Intuitively, higher efficiencies may allow for improved and faster transport of fluid and nutrients throughout the network, which are desirable capabilities for both vasculature and fungi. In general, it is expected that increases in the number of connections between the same set of nodes will improve network efficiency by creating shortcuts between pairs of otherwise more distant regions. But an addition of edges will in turn increase the material cost of the network. Here, we explicitly define the cost to be the sum of the physical lengths of all edges in the system (see ~\ref{s:methods}). As noted in the discussion section below, an improved quantification of network cost would include radial information as well, but this data was unavailable for the vasculature systems.
To quantify the balance between competing network features, we assess the tradeoffs between the \textit{relative} network efficiency ($E_{\text{rel}}$) and wiring ($W_{\text{rel}}$) by considering quantities that reflect how close a given network is to spatially informed null models \cite{Buhl:2004a,Buhl:2006a,Cardillo:2006a}. The two null models are the minimum spanning tree \textit{MST} (Fig.~\ref{f:null_models}A) and the greedy triangulation \textit{GT} (Fig.~\ref{f:null_models}B), which are good approximations for the least efficient and costly, and for the most efficient and costly planar graphs, respectively \cite{Buhl:2004a}. The definitions of $E_{\text{rel}}$ and $W_{\text{rel}}$ are bounded between 0 and 1 to reflect the similarity of the real networks to their \textit{MST} and \textit{GT}, respectively (see \textit{Methods}, \textit{S.I.}). We observe that all networks lie below half of the maximum value of the relative material cost ($W_{\text{rel}} < 0.5$), suggesting the presence of metabolic constraints on the geometric structure of these systems (Fig.~\ref{f:wiring_eff_paths}B). In addition, these constraints appear stronger for the vasculature networks (with $W_{\text{rel}}$ ranging from approximately 0.05 to 0.1) than for the fungi (with $W_{\text{rel}}$ ranging from approximately 0.2 to 0.4), suggesting a different set of pressures and optimization principles in the two organisms.

Notably, the vasculature consistently satisfies $W_{\text{rel}} < E_{\text{rel}}$, with a significant margin between the two quantities. This means that for a given cost, these networks achieve a relatively greater efficiency, implying an economical, well-organized architecture. A similar finding holds for several of the fungi, but we also find that some of the mycelial networks lie just above, on, or sometimes below the line of equality ($W_{\text{rel}} = E_{\text{rel}}$), suggesting a less-optimal use of extra material as measured by its ability to increase the routing capabilities of the network.  Moreover, for significantly lower $W_{\text{rel}}$, the rat vasculature achieves similar or higher $E_{\text{rel}}$ values than many of the fungi, indicating consistent and advantageous use of material. Finally, we note that the rodent data is much more tightly clustered in this phase space compared to the more varied fungal data; this points to consistent, regularized network structure in the vasculature in contrast to more diversity in the mycelial networks.

\subsection{Differences in network robustness}

To better understand the implications of the wiring-efficiency tradeoff, we also considered network robustness, which -- like efficiency -- should be facilitated by increased wiring. We let the robustness, $R$, of a network be the percentage of edges removed in order for the size of the largest connected component to drop to half of its original value \cite{Buhl:2004a,Buhl:2006a}, and we consider the \textit{relative robustness} $R_{\text{rel}}$ (\textit{Methods}, \textit{S.I.}). We observe a clear, positive trend between $R_{\text{rel}}$ and $W_{\text{rel}}$ across organisms and species (Fig.~\ref{f:wiring_eff_paths}C), indicating that increasing the amount of material used to connect a given set of nodes strongly correlates with improved resistance to network damage. This was not as clear in $W_{\text{rel}}$ \emph{vs.} $E_{\text{rel}}$, suggesting that improved efficiency is not always solely a result of more wiring, but is also determined by the specific placement of that material in the network. Notably, all data lies on or above the line $R_{\text{rel}} = W_{\text{rel}}$, meaning that these biological systems are able to achieve higher robustness with an equal or relatively smaller amount of wiring. There is also a large separation between the fungal networks and the vasculature, with all of the fungal networks exhibiting greater $R_{\text{rel}}$, suggesting that they are better optimized for resistance to damage. These findings point to the fact that although comparable in certain ways, the two different types of transport networks indeed exhibit different strengths.

\begin{figure}[h]
\includegraphics[width = \linewidth]{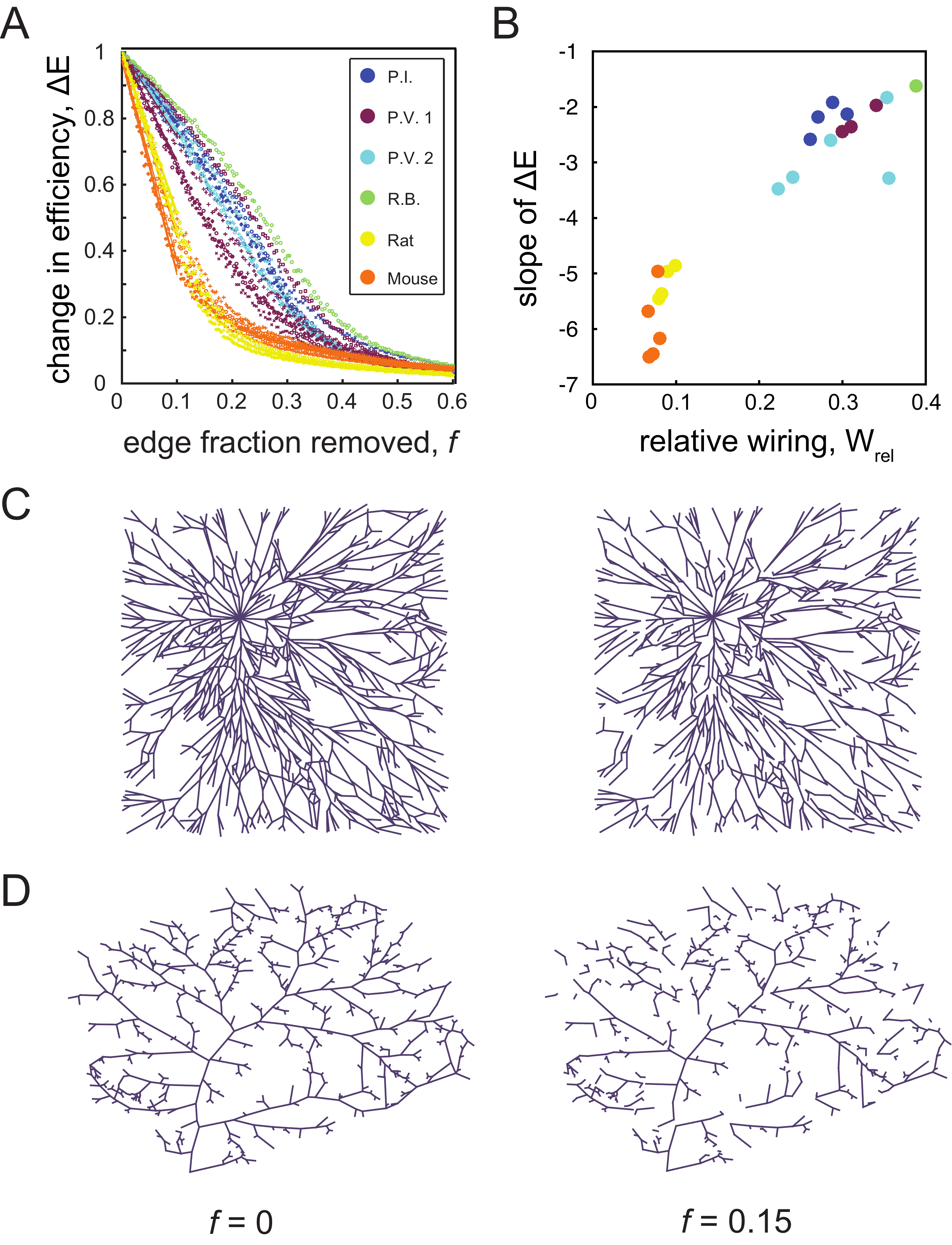}
\caption{\textbf{Change in network efficiency with edge removal.} \emph{(A)} The change in global efficiency as a function of edge fraction removed, showing that the efficiency of the vasculature networks declines more rapidly than that of the fungal networks. \emph{(B)} The slope of the linear drop-off in efficiency is computed for each network, and then plotted against the relative wiring. \emph{(C)} An example of a fully intact network from \emph{P.V. 2}, and the same network after a fraction $f = 0.15$ of edges have been removed at random. \emph{(D)} An example of a fully intact network of a mouse vasculature system, and the same network after a fraction $f = 0.15$ of edges have been removed at random.}
\label{f:edgeRemoval_Efficiency}
\end{figure}

Given that both the vasculature and fungal networks are subject to certain forms of impairment, another important consideration is how the efficiency of the network changes as a result of damage. In other words, how functional does the topology of the network remain when edges are removed from the system? In order to examine this question, we tracked the change in the efficiency of each network as a function of edge fraction removed. Specifically, if $f$ is the edge fraction removed, we compute $E(f)/E(0)$ for a range of $f$, where $E$ is the global efficiency of the real networks (see ~\ref{s:methods}). The structure of the resulting curves provides insight into how the transport capabilities of each network are altered under damage (Fig.~\ref{f:edgeRemoval_Efficiency}A; see~\ref{s:methods}). We observe that the efficiency of the vasculature networks falls off more rapidly than the efficiency of any of the fungal networks.

In order to help visualize the effect of the edge removal, we show an example of a robust fungal network (Fig.~\ref{f:edgeRemoval_Efficiency}C) and more delicate vasculature network (Fig.~\ref{f:edgeRemoval_Efficiency}D) with no edge removals ($f = 0$) and after a fraction $f = 0.15$ edges have been removed. One observes that while the fungal network maintains a large connected components after edge removal, the vasculature system has become highly disjoint. To quantify the rates of decline in efficiency for realistic removal fractions, we considered the slope of linear fits of each curve between $f = 0$ and $f = 0.1$. Then, we asked whether the steepness of the fall-off was related to the relative wiring of the network. Turning to Fig.~\ref{f:edgeRemoval_Efficiency}B, we find that $W_{rel}$ is indeed a good predictor of the decline in function (correlation $r = 0.946$, $p$-value $= 2.853e-11$). This analysis highlights the fragility of the vasculature system, whose ability to maintain high performance becomes impaired quickly with damage compared to the fungal networks, which appear better able to maintain efficiency in the face of edge losses.
\section{Discussion}
\label{s:discussion}
Much of network science revolves around uncovering common organizational principles that exist across many different kinds of networks. For example, a wide variety of networked systems -- both natural and man-made -- have been shown to display self-similarity, often manifesting in power-law degree distributions \cite{Albert:2002a}. In spatial networks, power laws are less expected in the degree distribution, but they have been observed in edge length and betweenness centrality distributions of cities, and in scalings of edge weights, distances, and centrality with degree \cite{Barthelemy:2011}, to name a few examples. It has also been suggested that biological distribution networks, including blood vasculature and respiratory systems, are fractal-like and space filling \cite{West:2005a, Boddy:2008a}, and these properties have been used to validate and explain the 3/4 scaling laws that relate metabolic rates to body mass \cite{West:1999a}. More recently, hierarchical structure in the form of nested loops has been studied in the distribution networks of leaf veins \cite{Katifori:2012a}, and this structure has been shown to arise in networks optimized for resistance to damage and fluctuations in load \cite{Katifori:2010a,Corson:2010a,Ronellenfitsch:2016a}. Providing quantitative measures of what it means to have hierarchical organization is thus an important line of inquiry for spatially embedded and biological networks.
\subsection{Rentian scaling as a common design principle}
Here, we showed that both fungal and vasculature networks contain a form of hierarchical structure in physical and topological space, as evidenced by Rentian scaling. We then characterized the complexity of that network architecture by comparing the physical and topological scaling exponents, finding that these distribution systems are cost-efficiently embedded into Euclidean space, a feature also exhibited by the \emph{C. Elegans} connectome, human structural brain networks and computer circuits \cite{Bassett:2010a}, and the London Tube transportation system \cite{Sperry:2016a}. In biological transport networks, the presence of Rentian scaling provides robust evidence of common emergent design principles that persist across species and organisms, and notably, of organization that has arisen through natural evolution in the absence of global rules for network construction. Physical Rentian scaling in particular is indicative of self-similar architecture over different length scales in real space, and because the scaling relationship is determined from a random sampling of the network, its existence signifies a degree of spatial homogeneity and space-filling structure. Biological transport networks have previously been compared to man-made transit systems \cite{Tero:2010a}, and here we also find that quantitatively, the observed scaling relationships of the biological systems are most similar to those observed in the London tube network ($p = 0.4$ and $t = 0.31$) \cite{Sperry:2016a} with low scaling exponents indicative of highly constrained, planar (or near-planar) topology, and are least similar to the \emph{C. Elegans} connectome and human structural brain networks ($p =  0.74, 0.78$, respectively, and $t = 0.77, 0.78$, respectively) \cite{Bassett:2010a,Klimm:2014a}, which exist in 3-dimensional space and exhibit more long-distance and crossing edges \cite{Bullmore:2009aa,Bassett:2010a}.

In addition to characterizing the empirical data, an important question is what the origins of such observed scaling principles might be. To gain further insight into how this hierarchical structure could arise, we used a resistor network framework to generate a collection of model networks optimized for transport efficiency in the presence of fluctuations in load. Though the model used here is not constructed to imitate the specific details of each network, it indeed captures several features of the real data that are likely crucial for the physical and topological scaling and the relationships between them. In particular, the model takes into account constraints on network ``cost", produces networks with some redundancies, and yields space-filling architectures. In combination, these properties give rise to the observed hierarchical organization in physical and topological space. Note that optimization without fluctuations tends to produce trees as the optimal networks \cite{Durand:2007a, Bohn:2007a}, which could still yield Rentian scaling in physical space, but would lack topological scaling. Furthermore, without the space-filling nature seen in the model networks, one could still uncover topological scaling, but large spatial inhomogeneities in either the distribution of nodes or edges would destroy the scaling in real space, as observed in a minimally-wired version of the \emph{C. Elegans} neuronal network \cite{Bassett:2010a}.

\subsection{Topo-physical constraints on biological distribution networks}
In general, we find less variation in the physical (\emph{vs.} topological) scaling exponents, suggesting constrained physical layout across organisms. However, the fungi and vasculature do seem to exhibit some differences in their positioning in the topo-physical phase space. Vasculature systems are consistently closest to the theoretical minimum Rent's exponent, and on the low end of \textit{both} the topological and physical scaling spectra, suggesting that they may be especially efficiently wired for low-cost and short node-to-node connections. We also find that while the physical scaling exponents for the vasculature are tightly clustered, the fungal networks exhibit more spread along this dimension across species, indicating greater variation in the spatial structure and complexity of their networks. We hypothesize that these differences may be attributable to the highly controlled \emph{vs.} more heterogeneous environmental conditions of vasculature and fungi, respectively. In particular, the mycelial networks are not required to service a bounded region of space, and the embedding space itself (i.e., the soil and availability of water and nutrients) can also be heterogeneous and impact network development \cite{Ritz:2004a}. These conditions may lead the fungi to grow and adapt in more varied and unconstrained patterns that differ within and across species.

\subsection{Biological distribution networks display tradeoffs in cost, efficiency, and robustness}
In the second part of the analysis, we set out to quantify the relationships between three desirable but competing features of distribution systems - material cost, transport efficiency, and robustness - finding clear distinctions in how the vasculature and fungi balance these quantities. We hypothesize that the differential tradeoffs observed in fungal \emph{vs.} vasculature networks may reflect their diverse functions and environmental settings. To be effective, the vasculature network must supply a fixed area with blood, and thus experience a notion of boundary conditions and minimum requirements that may constrain the possibilities or necessity for further growth. In contrast, the growth of mycelial networks need not be constrained such that they occupy a pre-allocated region of space. One hypothesis for the near-minimal wiring in the vasculature may be that it is energetically \textit{necessary}, and another reason might be that it is \textit{sustainable} due to the confined and protected setting of the skull. Perhaps the vasculature can afford to devote less material to the network because it resides in a regulated setting, whereas the mycelial systems typically do not. In particular, environmental changes and patchy habitats are likely to play a large role in the supply of nutrients to the fungi, but the supply of oxygenated blood through the carotids to the brain is well regulated and near constant; thus, we might expect less load variation in the vasculature compared to the fungal systems. Importantly, these differences impact the networks' efficiencies and robustness. The vasculature systems are able to achieve relatively high efficiency with low cost, suggesting that the network architecture is structured in order to utilize a small amount of wiring redundancy in a way that significantly improves the capabilities of the network in terms of its ability to route resources. In contrast, the fungal networks strike a different balance, and tend to be more expensive and not always consistently organized such that higher relative material costs directly yield relatively greater relative efficiencies. This indicates less standardized and perhaps less optimized network architectures. Notably, however, the tradeoff of greater material cost in the mycelial systems directly yields improved resistance to damage, and the ability to maintain higher functionality under perturbation of the network. This feature may be demanded for survival or driven by the more varied and unprotected environmental conditions in which these systems reside.

\subsection{Methodological considerations and future directions \label{s:considerations}}

In this investigation, we utilized a network-theoretic representation and analysis to gain insight into the commonalities and distinctions in the organization of two different biological transport systems. In the first portion of the study, we considered the (unweighted) binary graph structure and showed that the vasculature and fungi both exhibit fractal-like architecture in the form of Rentian scaling. In the second part, we also considered weighted versions of each network, where edges were assigned a value equal to their length in Euclidean space. This analysis allowed us to estimate the material costs and efficiencies of the distribution systems. It is crucial to point out, though, that we did not use information about cord or vessel radii, the main reason being that this information was unavailable for the vasculature. However, it is known that tube area has important affects on transport in distribution networks, since it is related to flow resistance. Including this additional physical property would thus yield a more accurate analysis, both in terms of quantifying network costs as well as measuring functional properties of the network such as efficiency. Indeed, past studies on mycelial networks have found that the presence of radial thicknesses confers improved transport characteristics to the system \cite{Bebber:2007a}. It would be interesting to take this into account in future work to investigate whether the inclusion of radial information in the weighting of the network reduces or enhances the differences found between the vasculature and fungal systems. Finally, we also note that network efficiency quantifies flow only in terms of shortest paths, and in this case studied here, assumes bi-directionality of flow; an understanding of transport along indirect walks is also relevant in distribution systems, as is the notion of directed transport that allows for the movement of nutrients through long distances. The extension of traditional network measures and null models to include this type of information may lead to considerable insight into the structure/function relationship in biological transport networks more generally.

Aside from those already mentioned, there are several other directions for future work. One interesting investigation would be to see if Rentian scaling also appears in other transport networks, for example, in the vasculature systems of larger organisms and humans. On the other hand, one could also try to further understand the observed differences between the structure of the vasculature and fungi. Here, we suggested that they may be due to the different environmental habitats and function of the two systems, but to establish this will likely require both further experimental work, as well as more extensive theoretical models. For example, it would be interesting to build off of prior work \cite{Bohn:2007a,Hu:2013a,Katifori:2012a,Ronellenfitsch:2016a} and continue to construct models that describe how a distribution network evolves in a physically un-bounded and dynamically variable environment, \emph{vs.} in a spatially constrained and more regulated environment, and investigate how these differences affect network adaptation and the resulting network architecture. Another intriguing direction would be to further investigate the embedding of distribution networks that live not in 2-dimensions, but in 3-dimensions \cite{Blinder:2013aa, Modes:2016a}. While there are often strong correlations between topology and geometric structure in planar transport networks, these relationships may be more complex and rich for networks that exist in three spatial dimensions. For example, one could inquire about the values and relationships between physical and topological exponents in this higher dimensional space.
\subsection{Conclusion}
Network science has provided a powerful set of tools to untangle complex systems and find emergent and common organizational principles. Using a network science based analysis of two distinct biological distribution networks, we have demonstrated conserved properties of efficient embedding and architectural complexity in both brain vasculature and fungi in the form of physical and topological Rentian scaling. Furthermore, we have uncovered fundamental differences in the topo-physical architecture of these systems that could be explained by differences in their function and environment. Our approach offers generalizable tools to quantitatively compare physical and topological network configurations in spatially embedded networks more broadly.
\subsection{Materials and Methods}
\label{s:methods}
\subsection{Data}
Rodent vasculature networks were kindly provided by Pablo Blinder and the group of David Kleinfeld at the University of California, San Diego. Further information about these networks can be found in \cite{Blinder:2010a}. All fungal networks were obtained from a freely available database at \cite{fungalData}; both the adjacency matrices and node coordinates were provided in the datasets. For a clean comparison against the vasculature, we studied a subset corresponding to mature, un-grazed networks grown without additional resources. In addition, we only considered networks with at least 500 nodes so as to obtain good estimates of the physical and topological Rentian scaling.

\subsection{Null models}
The two spatial null models examined were determined from the set of nodes and node locations of the real networks. The minimum spanning tree was computed on the Euclidean distance matrix between all node pairs, $D_{ij}$, using the algorithm from the MATLAB Boost Graph Library package \cite{matlabBGL}. The greedy triangulation was computed by connecting nodes in ascending order of the distance between them, while maintaining that no edges cross \cite{Buhl:2004a, Buhl:2006a, Cardillo:2006a}. Further details on these null models can be found in the \emph{S.I.}

\subsection{Network measures}
Given an Euclidean distance matrix $D_{ij}$ and an adjacency matrix $A_{ij}$ that describes node connectivity, the total wiring (i.e., material cost) of a network is given by
\begin{equation}
W = \sum_{i > j} A_{ij} D_{ij}.
\end{equation}
The average efficiency of a network is
\begin{equation}
E_{\text{avg}} = \frac{1}{N(N-1)}\sum_{i,j} \frac{1}{l_{ij}},
\end{equation}
where $l_{ij}$ is the shortest physical path between nodes $i$ and $j$, and the global efficiency $E$ is computed via normalizing $E_{\text{avg}}$ by the corresponding value for the fully-connected network with the same number of nodes \cite{Latora:2001a}. Following \cite{Buhl:2004a, Buhl:2006a, Cardillo:2006a}, the relative cost of a network was then defined as
\begin{equation}
W_{\text{rel}} = \frac{W - W_{\text{MST}}}{W_{\text{GT}} - W_{\text{MST}}},
\end{equation}
where $W$, $W_{\text{MST}}$, and $W_{\text{GT}}$ denote the total wiring of the real, \textit{MST}, and \textit{GT}, respectively. In a similar manner, the relative global efficiency, $E_{\text{rel}}$, is given by
\begin{equation}
E_{\text{rel}} = \frac{E - E_{\text{MST}}}{E_{\text{GT}} - E_{\text{MST}}}.
\end{equation}
Network robustness was probed by iteratively removing edges from the network at random, and tracking the size of the largest connected component after each removal. This process was repeated 20 times for each network and we report averages over this ensemble. The relative robustness, $R_{\text{rel}}$, is then given by
\begin{equation}
R_{\text{rel}} = \frac{R - R_{\text{MST}}}{R_{\text{GT}} - R_{\text{MST}}}.
\end{equation}
Each of the relative measures are normalized between 0 and 1 in order to reflect the closeness of the real networks to their corresponding \textit{MST} and \textit{GT}, respectively.

For the random edge removal \emph{vs.} change in efficiency analysis, we tested removal fractions between $f = 0$ to $f = 0.6$, in steps of $\Delta f = 0.004$. At each step, we removed the corresponding fraction of edges at random from each network, and then computed the resulting global efficiency of the perturbed system. This process was repeated 20 times for each fraction $f$. The curves in Fig.~\ref{f:edgeRemoval_Efficiency} depict the value of $E(f)/E(0)$ averaged over this ensemble.

\section*{Acknowledgements}

LP acknowledges support from the NSF Graduate Research Fellowship Program. DSB and LP also acknowledge support from the John D. and Catherine T. MacArthur Foundation, the Alfred P. Sloan Foundation, the Army Research Laboratory and the Army Research Office through contract numbers W911NF-10-2-0022 and W911NF-14-1-0679, the National Institute of Health (2R01-DC-009209-11, 1R01HD086888-01, R01-MH107235, R01-MH107703, and R21-M106799), the Office of Naval Research, and the National Science Foundation (BCS-1441502, PHY-1554488, and BCS-1631550). DK acknowledges support from grant numbers R01-EB003832, R35-NS097265, and R01-MH111438. EK and HR acknowledge support from the NSF Award No. PHY-1554887 and the Burroughs Wellcome Career Award. PB acknowledges support from the European Research Council grant ERC-2014-STG-639416 and Israeli Science Foundation grant 1019/15.   FK acknowledges financial support from the EPSRC and MRC under gran number EP/L016044/1 and further contribution by \emph{e-Therapeutics  Plc.} The content is solely the responsibility of the authors and does not necessarily represent the official views of any of the funding agencies.



\clearpage
\newpage
\begin{flushleft}
\huge{\bf{Supporting Information}}
\end{flushleft}

\section{Details on network models}

\subsection{Fungal networks}
\label{s:fungal_networks}

\begin{figure*}
\begin{center}
\includegraphics[width = 5in]{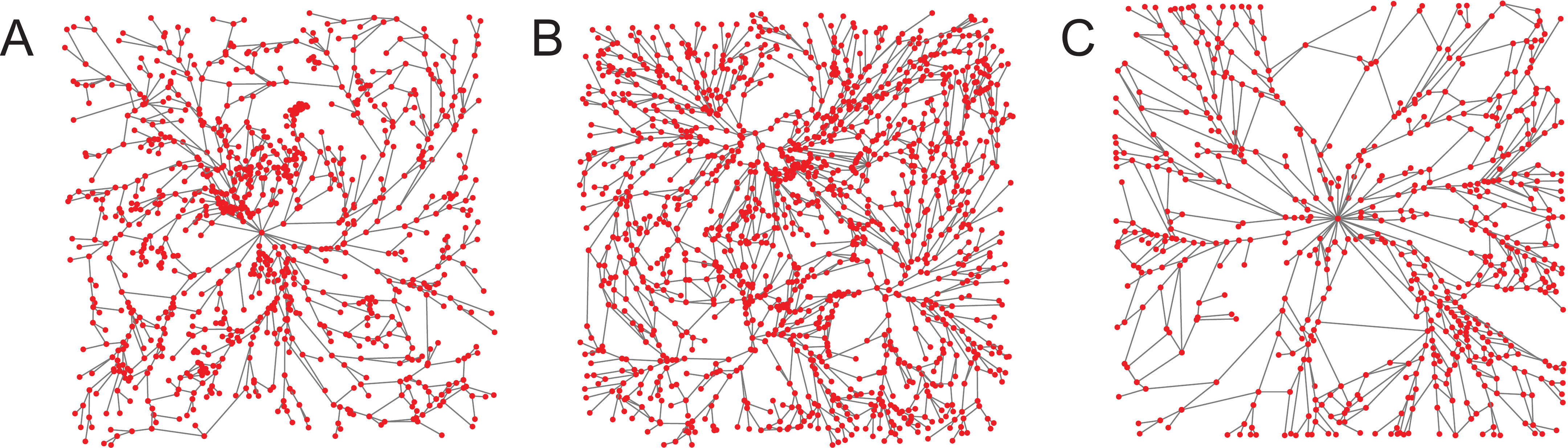}
\caption{\textbf{Additional examples of fungal networks.} \emph{(A)} The network structure formed by \textit{Phanerochaete velutina} grown from a single innoculum. \emph{(B)} The network structure formed by \textit{Phanerochaete velutina} grown from five innocula. \emph{(C)} The network structure formed by \textit{Resinicium bicolor} grown from a single innoculum.}
\label{f:fungal_SI}
\end{center}
\end{figure*}

Progress in image processing and analysis has allowed for the digitization and network extraction from images of fungi \cite{Heaton:2012a}. Using the typical graphical representation, the physical cords making up the organism are assigned to edges in the network, and branching, fusion, or end points among those cords are represented as nodes in the network. This construction results in a 2-dimensional, planar graph. In the binary representation, the adjacency matrix is given by

\begin{equation}
A_{ij} = \left\{ \begin{array}{ll}
         1 & \mbox{if there is an edge between nodes $i$ and $j$},\\
         0 & \mbox{otherwise} \end{array} \right.
\label{eq:binary}
\end{equation}

and captures topological information about the system. In general, the edges in a fungal network can also be described by a combination of their length and radius \cite{Heaton:2012a, Bebber:2007a, Fricker:2007a,Lee:2016a,Fricker:2008b}. However, since arterial thickness information was not available in the rodent vasculature, we only consider edge length in this study. In particular, given the set of spatial coordinates for each node $i$, $\{ x_{i}, y_{i}\}$, we compute the length of the edge from node $i$ to node $j$, $D_{ij}$, as the Euclidean distance between $i$ and $j$. This measure captures geographical information about how the nodes and edges are embedded into space, and allows one to estimate the material cost and efficiency of the network (see Sec.~\ref{s:network_measures}).

In this study, we examine three different species of mycelial fungi. For control in the comparison against the rodent vasculature, they are all un-grazed and grown without any additional resources aside from the source innocula. In addition, since the vasculature is sampled at only one time point, for each type of mycelium, we use the most mature (i.e., the oldest) networks. Finally, we only consider the subset of networks with at least 500 nodes, in order to obtain good estimates of the power-law Rentian scaling. In all, we examine 4 different networks formed by \textit{Phallus impudicus (P.I.)} grown from a single inoculum sampled at 46 days, 3 networks from \textit{Phanerochaete velutina (P.V. 1)} grown from 5 inocula and sampled between 30 to 35 days, 5 networks from \textit{Phanerochaete velutina (P.V. 2)} grown from 1 inocula and sampled between 39 to 46 days, and 1 network from \textit{Resinicium bicolor (R.B.)} grown from a single inoculum and sampled at 31 days. All of the data for the fungi (adjacency matrices and node locations) was collected from several studies and made freely available in a dataset that can be downloaded online \cite{fungalData}. More information on details of these networks can be found in \cite{Lee:2016a}, the references therein, and in the documentation within the database \cite{fungalData}.

We showed an example of the network formed by the fungus \textit{P.I.} in Fig.~\ref{f:networks}B of the main text. Here in Fig.~\ref{f:fungal_SI}, we also provide examples of \textit{P.V.2} (panel \emph{A}), \textit{P.V.1} (panel \emph{B}), and \textit{R.B.} (panel \emph{C}).

\subsection{Rodent vasculature networks}
\label{s:vasculature_networks}

\begin{figure}
\begin{center}
\includegraphics[width = 2.5in]{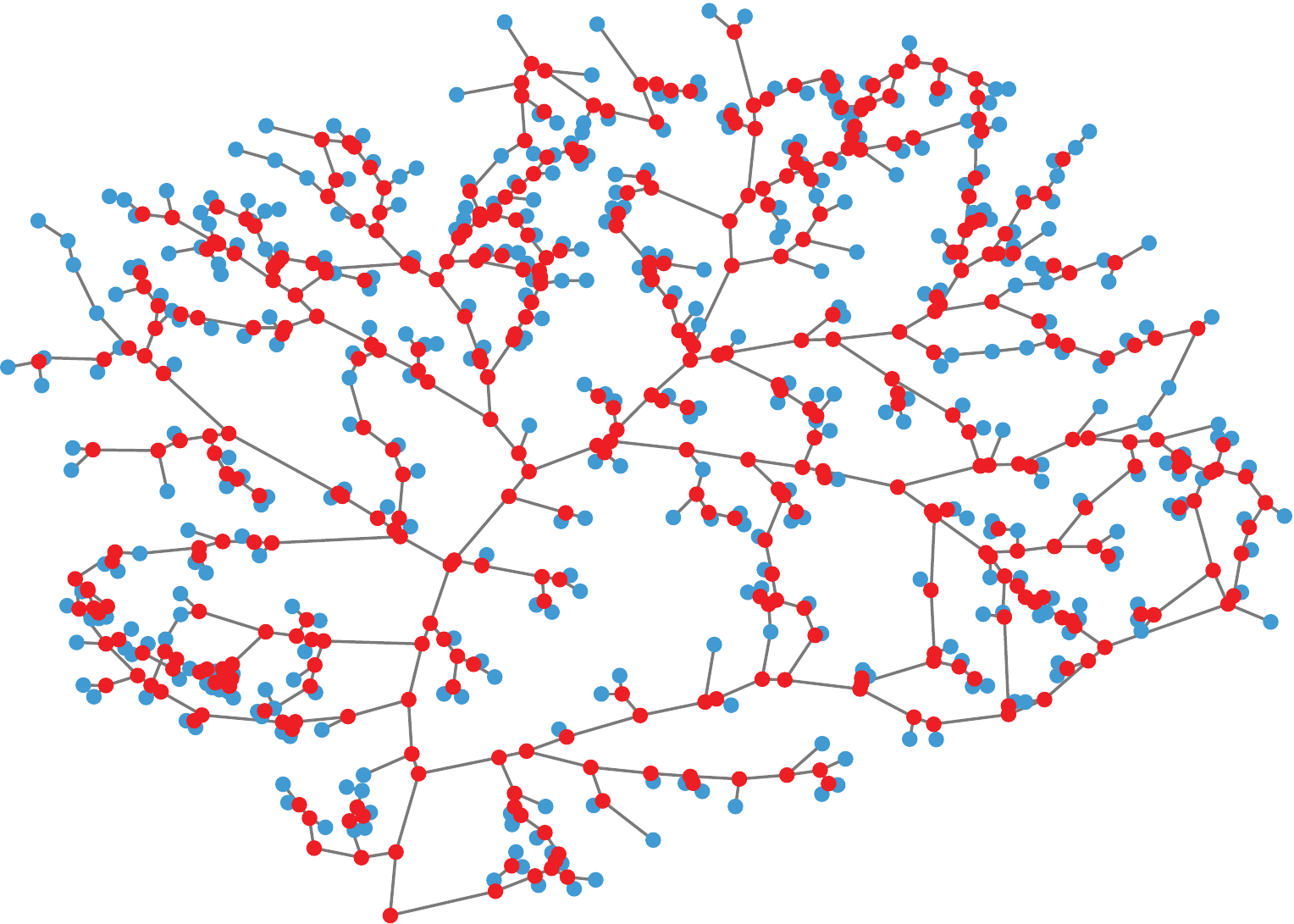}
\caption{\textbf{An example of the vasculature from mouse neocortex.} This figure shows the entire pial network from the territory of the middle cerebral artery in a mouse brain. In this network, nodes are either branching points (shown in red) or penetrating arterioles (shown in blue), and edges are vessel segments.}
\label{f:mouse_network}
\end{center}
\end{figure}

The second dataset consists of the pial networks in the region of the middle cerebral artery from four rats and five mice. To obtain a graphical representation, the vasculature was imaged and then traced by hand (further details can be found in the original study \cite{Blinder:2010a}). In this set of networks, nodes are either branching junctions where surface vessels come together, or penetrating arterioles that dive into and source the underlying microvasculature. Edges are vessel segments through which blood flows. As with the mycelial networks, the graphs describing the rodent vasculature are 2-dimensional and planar, and are also characterized by a set of node coordinates. Thus, these networks can be either binary (Eq.~\ref{eq:binary}) or weighted by edge length.

An example of the vasculature from a rat was shown in Fig.~\ref{f:networks}B; here we also include an example of the network from a mouse brain. Red nodes are branching points on the surface backbone and blue nodes are penetrating arterioles (Fig.~\ref{f:mouse_network}).

\section{Generating optimal transport networks}
\label{s:optimal_networks}

It has been shown in the past that many traits of organisms in general, and biological transport networks in particular, can be explained by the assumption that they are optimized for their function in some sense. In the case of transport networks such as leaf venation or mammalian capillary networks, one expects that they minimize the dissipated power in the network while at the same time being robust against external damage such as embolisms or vein occlusions~\cite{Katifori:2010a}. A developmental mechanism that can produce optimized networks using local adaptive rules was discussed in~\cite{Ronellenfitsch:2016a}.

In the following, we review the model of optimal transport networks from~\cite{Katifori:2010a} that is able to produce realistic efficient and robust network topologies.

Blood flow in the smallest capillaries can be modeled by laminar flow in cylindrical vessels of radius $R$ according to Poiseuille's law. Volume flow is then $Q = K \Delta p$, where the conductivity is $K = \frac{\pi}{8\mu L} R^4$, $L$ is vessel length, $\mu$ is fluid viscosity, and $\Delta p$ is the pressure drop along the vessel.

The capillary network itself is modeled as a graph with $N$ nodes where each edge $(ij)$ carries a volume flow $Q_{ij}$. At each node $i$ we have conservation of current,

\begin{align}
	\sum_j Q_{ij} = q_i, \label{eq:conservation}
\end{align}
where $q_i$ is the perfusion rate, or net current drawn from the network at the node. ~\eqref{eq:conservation} can be solved for the volume flows $Q_{ij}$ given the conductivities $K_{ij}$ and the net currents $q_i$.

An optimal network minimizes the dissipated power during operation, which is defined by

\begin{align}
	P = \sum_{(ij)} \frac{Q_{ij}^2}{K_{ij}}.
\end{align}
In order to force the network to be robust in the presence of fluctuating load,~\cite{Katifori:2010a} introduced
the load-averaged power dissipation,
\begin{align}
	\langle P \rangle = \sum_{(ij)} \frac{\langle Q_{ij}^2 \rangle}{K_{ij}}.
    \label{eq:averaged-pd}
\end{align}
Here, the angle brackets $\langle f \rangle = \frac{1}{N-1} \sum_{k} f^{(k)}$
denote an average over all possible network configurations
with net currents
\begin{align}
	q_i^{(k)} = \delta_{ii_0} - \delta_{ik}
\end{align}
where we fix the source $i_0$ and vary the sink $k \in \{1,\dots, N\} \setminus \{ i_0 \}$. This is called the moving sink model.

A well-posed optimization problem requires the introduction of a constraint that prevents the conductivities from diverging. A biologically sensible constraint is to set

\begin{align}
	\sum_{(ij)} K_{ij}^{\frac{1}{2}} = const.
\end{align}
This way, the total volume of the network is kept fixed, which can be interpreted as the organism having a finite amount of resources at its disposal to construct its venation network.

The resulting constrained optimization problem is then solved using a simulated annealing algorithm as in~\cite{Katifori:2010a}. As the initial condition we choose a triangular lattice network with randomly initialized  edge conductivities. The net currents $q_i$ are chosen to represent one source either at the center of the lattice or at one edge, and uniform sinks everywhere else, modeling uniform perfusion requirements.

After optimization, the result is a weighted network, where the weights correspond to the conductance between pairs of connected nodes. For each source location (edge or center), we constructed an ensemble of 50 simulated networks (100 in total). All conductances below a threshold value of 3.5e-8 were set to zero. Furthermore, since Rentian scaling only depends on the binary connectivity of a graph (and edge and node locations for physical scaling), after thresholding, we binarized each network by setting all remaining non-zero conductances to unity. This resulted in a set of unweighted adjacency matrices in the same form as the empirical data. Values of physical and topological scaling exponents reported in the main text correspond to an average over the 100 networks in the collection. Details on the Rentian scaling analysis are provided in the section below.

 \section{Rentian scaling analysis}

\subsection{Details on physical Rentian scaling}

As described in the main text, the physical Rentian scaling analysis is carried out by partitioning the networks into 5000 contiguous square boxes of different sizes in real space. The side length and center of each partition is chosen at random, but in order to avoid boundary effects due to the finite size of the network, we only sample boxes that are fully contained within the network's convex hull. For a given network, we compute the number of nodes $n$ inside each partition and the number of edges $e$ crossing the partition boundary, and plot them in $\log-\log$ space to assess if these quantities can be related by a power law of the form $e \propto n^{p}$, where $p$ is the physical Rent exponent. In order to robustly assess the data for this power law, we limited analysis to only include networks with at least 500 nodes.

\begin{table}[h]
\caption{The physical Rentian scaling exponents for the different classes of networks.}
\centering
\begin{tabular}{c c c c c c}
\hline\hline
Rat 	& 	Mouse 	& 	P.I. 	& 	P.V. 1 	& 	P.V. 2 	& R.B. 	\\
\hline
0.51 & 	0.53 		& 0.525	&  	0.542 	& 	0.538  	&  0.574 	\\
0.53 & 	0.52 		& 0.544 	&  	0.546	& 	0.543 	&  --- 	\\
0.52 & 	0.52 		& 0.534 	&  	0.570 	& 	0.539 	&  --- 	\\
0.52 & 	0.52 		& 0.514 	&  	0.583 	&	---    		&  ---		 \\
---    & 	0.53 		& ---	 	&  	0.571 	& 	---    		&  ---		 \\
---    & 	---    		& ---		&  	---       	& 	---    		&  ---		 \\
\hline
\end{tabular}
\label{t:pRent_p}
\end{table}

\begin{table}[h]
\caption{The physical Rentian scaling Pearson correlation coefficients.}
\centering
\begin{tabular}{c c c c c c}
\hline\hline
Rat 	    & 	Mouse 	& 	P.I. 	& 	P.V. 1 	& 	P.V. 2 	& R.B. 	\\
\hline
0.944   & 	0.912 	& 	0.956 	&  	0.952 	& 	0.951  	&  0.962 	\\
0.955   & 	0.934	& 	0.937	&  	0.945	& 	0.959	&  --- 	\\
0.953   & 	0.922 	& 	0.945 	&  	0.946 	& 	0.957 	&  --- 	\\
0.951   & 	0.926 	& 	0.953 	&  	0.943 	&	---    		&  ---		 \\
---    	   & 	0.930 	& 	--- 		&  	0.958 	& 	---     		&  ---		 \\
---    	   & 	---    		& 	--- 		&  	---       	& 	---    		&  ---		 \\
\hline
\end{tabular}
\label{t:pRent_r}
\end{table}

\begin{figure*}
\begin{center}
\includegraphics[width = 4.5in]{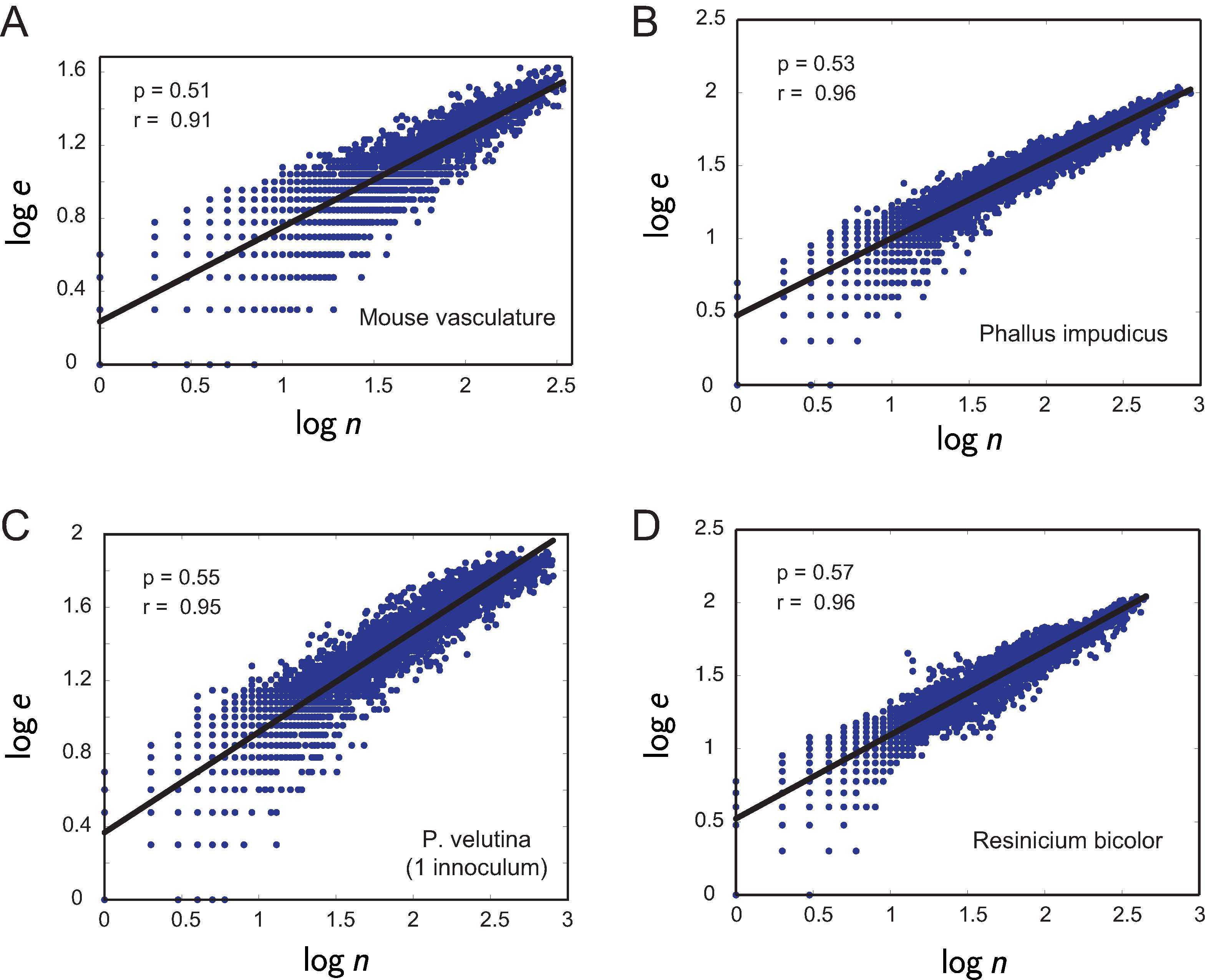}
\caption{\textbf{Examples of physical Rentian scaling in biological distribution networks.} To test for the presence of hierarchical structure in real space, we partition each network into boxes of different length scales, and count the number of nodes inside $n$ and the number of edges crossing the boundary $e$ of each box. Rentian scaling exists when these quantities scale with one another in $\log-\log$ space, and the physical Rent exponent $p$ is determined from the slope of the best fit line. \emph{(A)} An example of the physical Rentian scaling relationship from the arterial network of a mouse brain, \emph{(B)} from a network of the fungus \textit{P.I.}, \emph{(C)} from a network of the fungus \textit{P.V. 2}, and \emph{(D)} from a network of the fungus \textit{R.B.}. The lines correspond to a best fit line of the data points, from which we estimate the displayed scaling exponent $p$. We found that in all networks considered, there was a strong linear relationship (high Pearson correlation $r-value$) between $\log{e}$ and $\log{n}$.}
\label{f:pRent_SI}
\end{center}
\end{figure*}

In the main text, we showed examples of $\log{e}$ \emph{vs.} $\log{n}$ for one rat vasculature network and one network of the fungus \textit{P.V. 1}. Here, we also show examples of the scaling for a mouse vasculature network in Fig.~\ref{f:pRent_SI}A, and of the other types of fungi (\textit{P.I.}, \textit{P.V. 2}, and \textit{R.B.}) in Fig.~\ref{f:pRent_SI}B -- D. Plotting $e$ \emph{vs.} $n$ in $\log-\log$ space provides visual evidence of a power-law relationship, but to robustly quantify such a scaling, we used MATLAB's robust fit multilinear regression to estimate the physical scaling exponent $p$ from the slope of the best fit line through the data (displayed in the examples in Fig.~\ref{f:pRent_SI}). In addition, we used the value of the Pearson correlation coefficient $r$ to determine the goodness of fit. In Tables ~\ref{t:pRent_p} and ~\ref{t:pRent_r}, we give the physical scaling exponents for all networks, and the corresponding $r$-values for the linear regressions, respectively. As can be seen from the data, the scaling for the mouse vasculature appears to have the most spread, but all networks achieve high $r$-values $> 0.9$, and all fits were found to be statistically significant ($p$-values $< 0.05$).

An interesting feature of the rodent vasculature networks is the presence of ``end" nodes (Fig.~\ref{f:networks}), which are places where the surface backbone yields to form penetrating arterioles. Physically, this prevents a conservation of flux over the entire vasculature network, as compared to fungi. As found in \cite{Blinder:2010a}, most of the penetrating arterioles branch directly off the end of a surface vessel (as opposed to being located along a surface vessel or between two surface branching points). In order to investigate the effect of these end-point penetrating arterioles on the physical Rentian scaling relationship found in Sec.~\ref{s:pRent}, we iteratively removed the penetrating arterioles connected to only one other node in the network. This process yielded a reduced network on which we subsequently carried out the previously described physical Rentian scaling analysis. The exponents and corresponding correlation coefficients for the scaling relationships are shown in Table~\ref{t:pRent_reduced} and Table~\ref{t:pRent_corrReduced}, respectively.

\begin{table}[h]
\caption{The physical Rentian scaling exponents for the reduced rodent vasculature networks, in which the majority of penetrating arterioles have been removed.}
\centering
\begin{tabular}{c c}
\hline\hline
Rat 		& 	Mouse 	\\
\hline
0.491 	& 	0.506 	\\	
0.512 	& 	0.492 	\\	
0.504 	& 	0.506 	\\	
0.498   	& 	0.474 	\\	
---    		& 	0.506	\\
\hline
\end{tabular}
\label{t:pRent_reduced}
\end{table}

\begin{table}[h]
\caption{The physical Rentian scaling correlation coefficients for the reduced rodent vasculature networks, in which the majority of penetrating arterioles have been removed.}
\centering
\begin{tabular}{c c}
\hline\hline
Rat 		& 	Mouse 	\\
\hline
0.926	& 	0.905 	\\	
0.945 	& 	0.912	\\	
0.949 	& 	0.906 	\\	
0.944   	& 	0.895	\\	
---    		& 	0.909	\\
\hline
\end{tabular}
\label{t:pRent_corrReduced}
\end{table}

The reduced networks give qualitatively similar results compared to the full networks analyzed in the main text. In the case of the former, the scaling exponents $p$ are slightly smaller, but still close to 0.5 in all cases. We also observe decreases in the correlation coefficients for the scaling relationships, though this may be expected, since removing all of the end-nodes destroys some of the space-filling capacity of the network, and also disrupts spatial homogeneity.

\subsection{Details on topological Rentian scaling}

\begin{table*}
\caption{The topological Rentian scaling exponents.}
\centering
\begin{tabular}{c c c c c c}
\hline\hline
Rat 	    			& 		Mouse 			& 	P.I. 							& 		P.V. 2 	& 			P.V. 1		& 		R.B. 				\\
\hline
0.255 $\pm$ 0.001   		& 	0.203 $\pm$ 0.002	& 	0.317 $\pm$ 0.001 				&  	0.265 $\pm$ 0.001	& 	0.289 $\pm$ 0.001  	&  	0.325 $\pm$ 0.001 	\\
0.236 $\pm$ 9.337e-4  	& 	0.174 $\pm$ 0.001	& 	0.255 $\pm$ 0.001				&  	0.268 $\pm$ 0.001	& 	0.308 $\pm$ 0.001	& 	--- 				\\
0.252 $\pm$ 9.256e-4	& 	0.220 $\pm$ 0.001	& 	0.308 $\pm$ 0.001				&  	0.238 $\pm$ 0.002	& 	0.284 $\pm$ 0.001	&  	--- 				\\
0.232 $\pm$ 5.713e-4  	& 	0.124 $\pm$ 0.002 	& 	0.325 $\pm$ 0.001				&  	0.246 $\pm$ 0.002	&	---    				&  	---				 \\
---    	   				& 	0.177 $\pm$ 0.002	& 	---							&  	0.308 $\pm$ 0.001 	& 	---     			&  	---		 		\\
---    	   				& 	---    				& 	---							&  	---       			& 	---    				& 	 ---		 		\\
\hline
\end{tabular}
\label{t:tRent_t}
\end{table*}

\begin{table*}
\caption{The topological Rentian scaling Pearson correlation coefficients.}
\centering
\begin{tabular}{c c c c c c}
\hline\hline
Rat 		& 			Mouse 				& 	P.I. 	& 					P.V. 2 	& 					P.V. 1 	& 			R.B. 					\\
\hline
0.997 $\pm$ 2.125e-4  & 	0.993 $\pm$ 7.503e-4		& 0.999 $\pm$ 0.855e-4	&  	0.998 $\pm$ 1.028e-4  	& 		0.999 $\pm$ 1.075e-4  	&  0.998 $\pm$ 1.827e-4	\\
0.998 $\pm$ 1.769e-4  & 	0.993 $\pm$ 5.398e-4		& 0.995 $\pm$ 3.018e-4	&  	0.998 $\pm$ 2.161e-4 	& 		0.998 $\pm$ 1.152e-4 	&  --- 				\\
0.999 $\pm$ 8.617e-5  & 	0.994 $\pm$ 3.862e-4		& 0.999 $\pm$ 0.587e-4  	&  	0.998 $\pm$ 3.354e-4	& 		0.998 $\pm$ 1.296e-4 	&  --- 				\\
0.999 $\pm$ 8.076e-5  & 	0.993 $\pm$ 9.688e-4		& 0.999 $\pm$ 0.873e-4	&  	0.994 $\pm$ 5.733e-4	&		---    					&  ---					\\
---      			    & 	0.994 $\pm$ 8.834e-4 		& ---   				&	0.999 $\pm$ 1.397e-4	& 		---    					&  ---					 \\
---      			    & 	---    			  			& ---					&  	---       				& 		---    					&  ---		 			\\
\hline
\end{tabular}
\label{t:tRent_r}
\end{table*}

To test for the presence of topological Rentian scaling in the biological distribution networks, we employ a min-cut bi-partitioning algorithm. In particular, we utilize version 1.5 of the hyper-graph partitioning package \textit{hMETIS} \cite{Karypis:1999a, hMetis}, in combination with in-house MATLAB scripts. The software \textit{hMetis} recursively sections the network into halves, quarters, etc. (Fig.~\ref{f:tRent}A of the main text), in a way that attempts to minimize the number of edges passing from one partition to another. It is important to specify that unlike in the physical Rentian scaling analysis, these partitions do \textit{not} correspond to regions of the network in real space, but rather, are purely topological. After each round of partitioning, we track the number of nodes $n$ in each partition, and the number of edges $e$ crossing the partition boundary; if topological scaling exists, then plotting $e$ \emph{vs.} $n$ in $\log - \log$ space will yield an approximately linear relationship, and the slope of the line gives the topological Rentian scaling exponent $t$. In order to robustly assess the data for this power law, we limited analysis to only include networks with at least 500 nodes.

\begin{figure}
\begin{center}
\includegraphics[width = 2.5in]{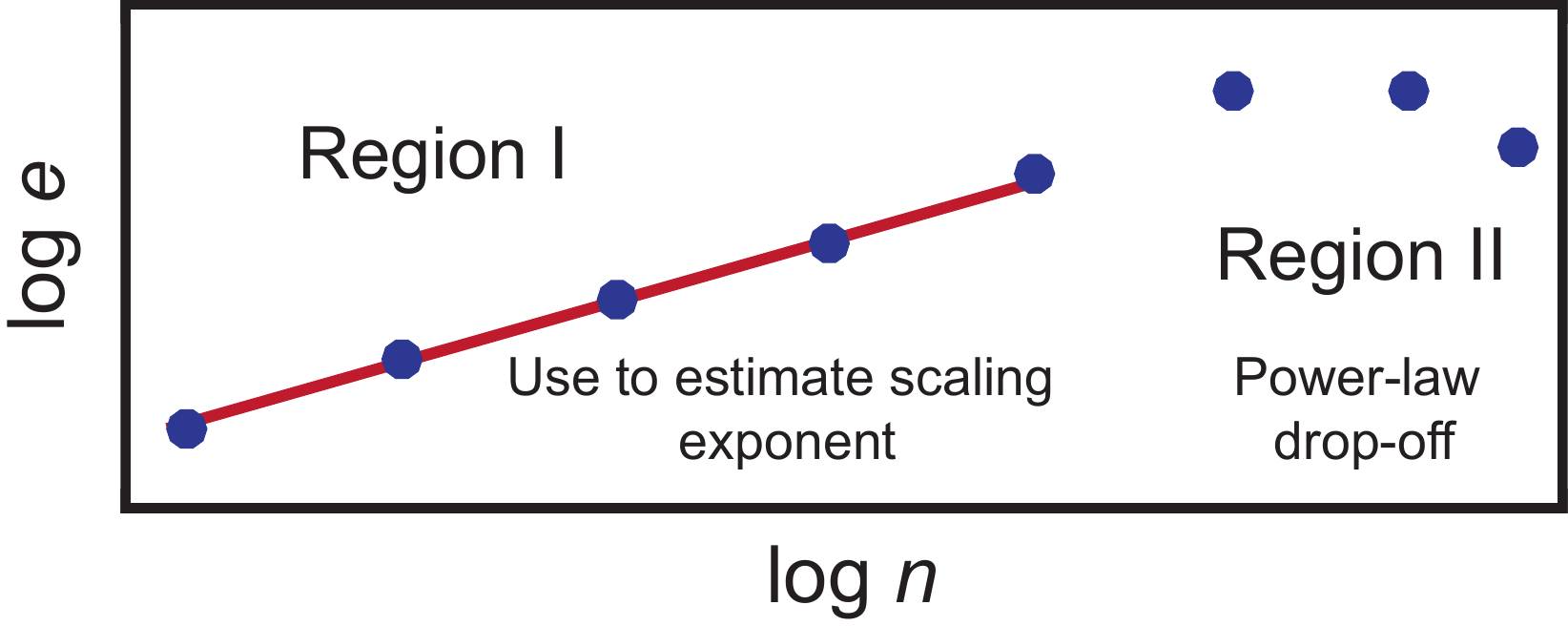}
\caption{\textbf{A schematic of ``Region I'' and ``Region II'' behavior in Rentian scaling.} Since the networks have a finite size, large topological partitions are subject to boundary effects that cause a drop-off in the power-law scaling (``Region II''). Therefore, in order to assess the network for fractal topology, we neglect the largest three partitions, and use the data in ``Region I''. If scaling exists, these points exhibit a linear relationship, and the topological exponent is determined from the slope of the line.}
\label{f:region2}
\end{center}
\end{figure}

An important point of consideration in this analysis is the occurrence of boundary effects due to large partitions. In the physical partitioning, boxes placed close to the boundary of the network will contain significantly fewer outgoing edges than boxes placed within the bulk, and so to obtain a good estimate of the scaling throughout the majority of the network, one should be cognizant of the finite size of the system (for example, we only include boxes within the network's convex hull). A similar issue arises in the topological scaling, where large partitions (i.e., on the same scale as the network as a whole) can exhibit large reductions in the number of crossing edges relative to the number of nodes inside, thereby skewing the topological scaling relationship. In the VLSI literature, this drop-off in the power law relationship between $n$ and $e$ is termed ``Region II'' \cite{Greenfield:2010a,Christie:2000a} (see Fig.~\ref{f:region2}). In line with common practice, we thus estimate the scaling exponent $t$ from ``Region I'', where the partition sizes are small enough to not be biased by boundary effects. For consistency across all networks, in each case, we neglected the first three sectioning steps (see Fig.~\ref{f:region2}).

An additional methodological consideration is that \textit{hMETIS} is not guaranteed to converge to an optimal solution, and therefore different runs of the algorithm (for the same network) may yield slightly different results. In order to take this variation into account, we ran the partitioning 50 times for each network. For each run, we then used MATLAB's robust fit multilinear regression to estimate the topological scaling exponent from the slope of the best fit line of $\log{e}$ \emph{vs.} $\log{n}$, and we used the values of the Pearson correlation coefficient $r$ to determine the goodness of fit. Here and in the main text, we report the average value of the exponents $t$ and correlations $r$ over the 50 trials, and the corresponding standard errors. We found that all networks considered exhibited strong topological Rentian scaling with high $r$-values (and $p$-values for all linear fits were statistically significant). In Tables ~\ref{t:pRent_p} and ~\ref{t:pRent_r}, we give the topological exponents for all networks, as well as the corresponding $r$-values from the linear regressions. As can be seen from the data, all networks achieve high $r$-values $> 0.98$.

\begin{figure*}
\begin{center}
\includegraphics[width = 5in]{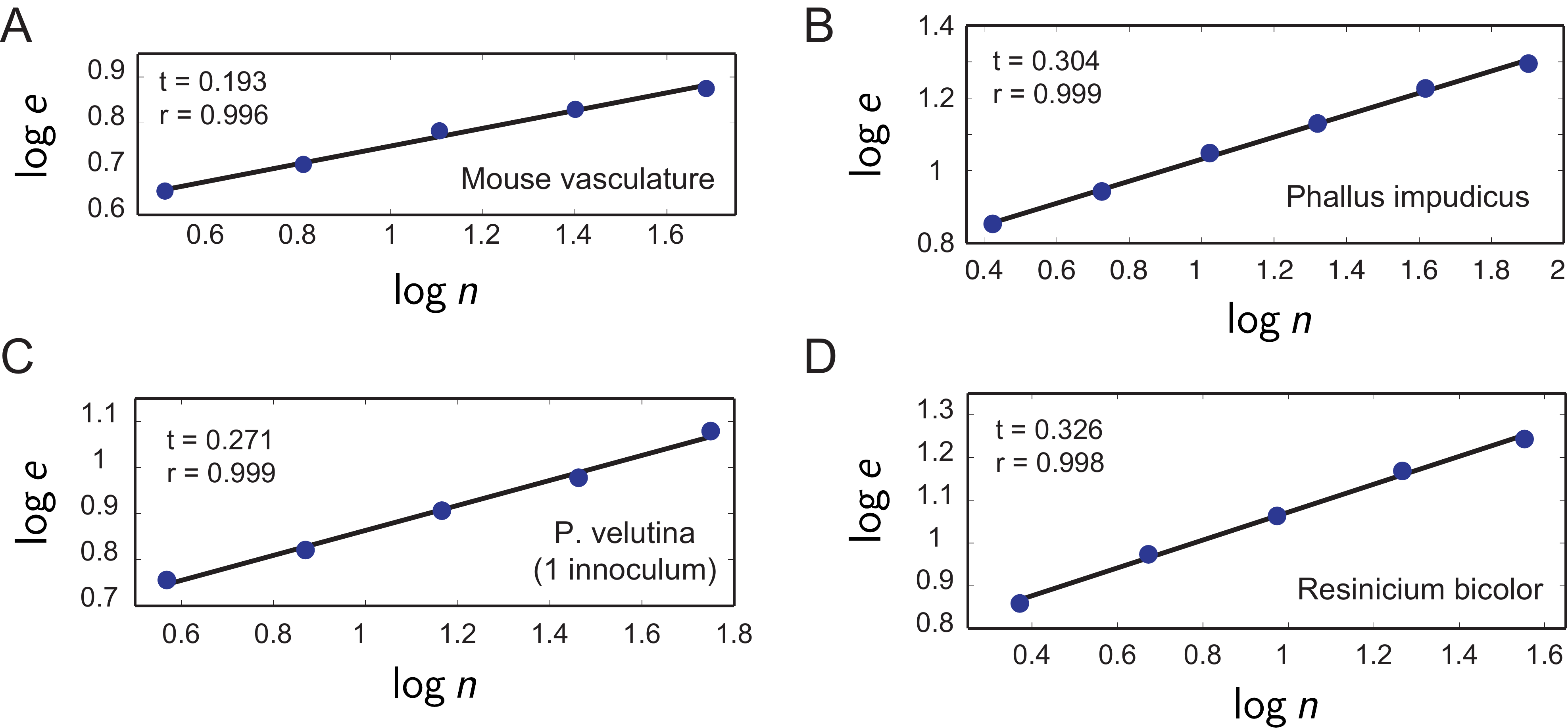}
\caption{\textbf{Examples of topological Rentian scaling in biological distribution networks.} To test for the presence of hierarchical structure in topological space, we recursively section each network into halves, quarters, etc., and count the number of nodes inside $n$ and the number of edges crossing the boundary $e$ of each partition. Rentian scaling exists when these quantities scale with one another in $\log-\log$ space, and the topological Rent exponent $t$ is determined from the slope of the best fit line. \emph{(A)} An example of the topological Rentian scaling relationship from the arterial network of a mouse brain, \emph{(B)} from a network of the fungus \textit{P.I.}, \emph{(C)} from a network of the fungus \textit{P.V. 2}, and \emph{(D)} from a network of the fungus \textit{R.B.}. The lines correspond to a best fit line of the data points, from which we estimate the displayed scaling exponent $t$. We found that in all networks considered, there was a strong linear relationship (high Pearson correlation $r$-value) between $\log{e}$ and $\log{n}$.}
\label{f:tRent_SI}
\end{center}
\end{figure*}

In the main text, we showed examples of the topological scaling relationship for a rat vasculature network and a network of the fungus \textit{P.V. 1}. Here, we additionally provide examples from one each of the additional types of networks considered (mouse vasculature, and the \textit{P.I.}, \textit{P.V. 2}, and \textit{R.B.} fungal networks), so as to provide a visual support and supplement for the reader (see Fig.~\ref{f:tRent_SI}).

As with the physical scaling analysis, we also considered topological scaling in the reduced networks with the end-point penetrating arterioles removed. The results are displayed in Tables~\ref{t:tRent_reduced} and ~\ref{t:tRent_corrReduced}. Without the set of end point nodes, we observe a slight increase in the topological scaling exponents, though the general range of values remains qualitatively the same. It is possible that the edges removed to create the reduced networks were not part of the set of edges that were cut in the original partitioning process, perhaps because the removed edges are not part of loops. On the other hand, since the topological scaling exponent does depends on the total number of nodes in each partition, the additional reduction in the number of nodes could lead to an increase in the scaling exponent.

\begin{table}[h]
\caption{The topological Rentian scaling exponents for the reduced rodent vasculature networks, in which the majority of penetrating arterioles have been removed.}
\centering
\begin{tabular}{c c}
\hline\hline
Rat 		& 	Mouse 	\\
\hline
0.262 $\pm$ 9.653e-4	& 	0.221 $\pm$ 0.003	\\	
0.238 $\pm$ 7.803e-4	& 	0.198 $\pm$ 0.003 	\\	
0.272 $\pm$ 7.815e-4 	& 	0.274 $\pm$ 0.002 	\\	
0.248 $\pm$ 9.693e-4  	& 	0.135 $\pm$ 0.002	\\	
---    					& 	0.193 $\pm$ 0.004	\\
\hline
\end{tabular}
\label{t:tRent_reduced}
\end{table}

\begin{table}[h]
\caption{The topological Rentian scaling correlation coefficients for the reduced rodent vasculature networks, in which the majority of penetrating arterioles have been removed.}
\centering
\begin{tabular}{c c}
\hline\hline
Rat 		& 	Mouse 	\\
\hline
0.998 $\pm$ 2.266e-4	& 	0.988 $\pm$ 0.002		\\	
0.999  $\pm$ 1.499e-4	& 	0.992 $\pm$ 8.193e-4 	\\	
0.999 $\pm$ 6.960e-5 	& 	0.996 $\pm$ 3.066e-4	\\	
0.998  $\pm$ 1.918e-4  	& 	0.993 $\pm$ 9.674e-4	\\	
---    					& 	0.992 $\pm$ 0.001		\\
\hline
\end{tabular}
\label{t:tRent_corrReduced}
\end{table}


\subsection{Rentian scaling for idealized network topologies}

In order to better frame the scaling exponents of the real fungal and vasculature networks, we compared them to the exponents from a set of idealized topologies: triangular lattice, square lattice, hexagonal lattice, and minimum spanning tree. The values of $p$ and $t$ for each of these networks are reported in Table~\ref{t:rent_ideal} of the main text. Each lattice was constructed in a square region. For a proper comparison among the different types, we fixed the side of each unit (triangle, square, hexagon) to be equal to 1, and fixed the number of nodes to be approximately the same for each lattice and on the same order as the mean number of nodes across all of the real data ($\overline{N} = 1228$). (Due to the different symmetries of each lattice, a perfect matching of the number of nodes was not possible while simultaneously ensuring the lattices were contained in a square region with equal side length). We thus used $N = 1380$ for the triangular grid, $N = 1369$ for the square grid, and $N = 1373$ for the hexagonal grid, which we found to be the best balance between having a near constant number of nodes across the different lattices while also being close to the mean number of nodes across all of the real networks. As described above, the topological partitioning was run 50 times for each network, and we report average values of the topological scaling exponent. Physical and topological scaling was significant in all lattices, with Pearson's correlation coefficients for the scaling relationship satisfying $r > 0.9$ in each case.

For the minimum spanning tree (\textit{MST}), we placed $N = 1374$ nodes (the mean number of nodes used for the lattices) at random in a box of equal side length. The spanning tree was then computed on that set of points using the method described below, in Sec.~\ref{s:null_models}. Since different instantiations of the node placements yield different spanning trees, we constructed 50 networks -- each from different random initialization of the nodes -- and reported average values of the physical Rentian scaling exponent. In each case, the scaling was significant, with $r > 0.9$. The spanning trees do not exhibit topological scaling.

\section{Details on network null models}
\label{s:null_models}

In spatial networks, the nodes and edges exist in real Euclidean space, and this physical embedding often has significant consequences for the network topology \cite{Barthelemy:2011}. For example, both the fungal and vasculature networks are \textit{planar} (or nearly planar), meaning that no edges cross. An important constraint for biological systems, in particular, may be the material and energetic costs associated with building and maintaining the physical structure of the network. But in competition is the fact that distribution networks must be able to move resources efficiently and be robust to damage. In order to gain an understanding of how the physical embedding of these networks might affect their architecture, and how distinct types of biological transport systems may differentially balance these pressures, we examined two planar null model networks: the minimum spanning tree (\textit{MST}) and the greedy triangulation (\textit{GT}). The same or similar null models have previously been used in the analysis of fungal systems, \cite{Heaton:2012a, Bebber:2007a,Fricker:2008b,Fricker:2008c, Fricker:2009a}, slime mould \cite{Tero:2010a}, ant networks of galleries \cite{Buhl:2004a}, and urban street networks \cite{Buhl:2006a, Cardillo:2006a}.

The \textit{MST} is a planar graph that connects all of the nodes in a network such that the sum of the total edge weights is minimum. By construction, the \textit{MST} also contains the minimum number of edges $M$ needed to connect the network ($M = N - 1$, where $N$ is the number of nodes). In the biological distribution networks studied here, the relevant edge weights are their physical lengths. Therefore, we preserve the true geographic locations of all nodes in each network, and compute the \textit{MST} on the matrix of Euclidean distances, $D$, between all node pairs. This network minimizes the total material cost $W$ of the network (defined as the sum of all edge lengths):

\begin{equation}
W = \sum_{i > j} A_{ij} D_{ij},
\label{eq:total_wiring}
\end{equation}

and thus allows for testing the extent to which biological distributions systems are optimized for minimal wiring. We computed the \textit{MST}s for all networks using the algorithm from the MATLAB Boost Graph Library package \cite{matlabBGL}.

The \textit{GT} lies on the opposite end of the spectrum, and is a maximally connected -- in terms of number of edges -- planar network. In particular, following \cite{Buhl:2004a,Buhl:2006a,Cardillo:2006a}, we compute \textit{GT}s on the true node locations of all real networks by iteratively connecting pairs of nodes in ascending order of their distance, while maintaining that no edges be allowed to cross. As with the \textit{MST}, this null model is also constructed under a geographical constraint, but since it contains many more edges, it represents an effective upper bound on the material cost of a planar network. Thus, the \textit{GT} can be used to gain an understanding of how the structure and capabilities of the true vasculature and fungi compare and contrast to networks where the total wiring cost is not optimized for.

\section{Relative measures of cost, efficiency, and robustness}
\label{s:network_measures}

In this study, we investigate similarities and differences between two distinct types of biological distribution networks, in terms of how they are simultaneously optimized for cost, efficiency, and robustness. It is important to note that these measures have previously been used to quantify and compare the structure of fungal networks to low-cost and high-cost null models \cite{Heaton:2012a, Bebber:2007a,Fricker:2008b,Fricker:2008c, Fricker:2009a}. Here, we build on and extend prior analyses along this line of inquiry by examining the \textit{relative} tradeoffs and relationships between this set of quantities, and most importantly, by comparing and contrasting these tradeoffs across organisms. In particular, we utilize a set of metrics that measure how close or far a given real network is from its two extreme null models, allowing for a direct quantification and comparison of the vasculature and fungi. Below, we explain these quantities in more detail.

We first note that in spatial networks such as those considered here, it is often \textit{physical} rather than \textit{topological} distances that are important. Therefore, rather than the total number of edges, the material cost of a network is defined to be the sum of the lengths of all connections (Eq.~\ref{eq:total_wiring}), which provides an estimate of how ``expensive" the system is to construct and maintain.

Along a similar line of reasoning, the shortest physical path between any two pairs of nodes $i$ and $j$ is taken to be the path that minimizes the sum of the edge lengths between those nodes. This is in contrast to the shortest topological path between the same nodes, which is that which connects $i$ and $j$ along the fewest number of edges, and thus neglects geographic information. To illustrate the differences, Fig.~\ref{f:wiring_eff_paths}A of the main text shows a toy network that highlights the shortest physical path and topological path between nodes $i$ and $j$. Knowing the shortest paths between all node pairs allows for the calculation of the average efficiency \cite{Latora:2001a}, $E_{avg}$, which is one way to quantify the ease of flow and communication between nodes of a network. Denoting the length of the shortest physical path between $i$ and $j$ as $l_{ij}$, the average efficiency is given by

\begin{equation}
\label{eq:globalE}
E_{\text{avg}} = \frac{1}{N(N-1)}\sum_{i,j} \frac{1}{l_{ij}}.
\end{equation}

It is common practice to normalize the efficiency for a given network with $N$ nodes by its value for the corresponding fully-connected $N$-node network. This results in the \textit{global efficiency}, which we denote simply as $E$. Shortest path and efficiency measures have previously been used to study spatial networks such as ant galleries \cite{Buhl:2004a}, street patterns in cities \cite{Buhl:2006a,Cardillo:2006a}, the brain \cite{Bullmore:2009aa}, slime mould \cite{Tero:2010a}, and fungal networks \cite{Heaton:2012a, Bebber:2007a,Fricker:2008b,Fricker:2008c, Fricker:2009a}, but to the best of our knowledge, have not yet been used to study vasculature.

Another desirable feature for both natural and man-made networks is robustness. In other words, how well-connected does the system remain when subjected to damage? This is an important consideration in both the vasculature and fungal networks. In fungal networks, damage can occur either from rough environmental conditions or from predation \cite{Heaton:2012a,Boddy:1999b, Boddy:2007a,Rotheray:2008a}, and in vasculature, a common source of damage is from the formation of blood clots that block vessels and prevent flow, leading to stroke. Previous work on fungal networks has shown that these organisms can achieve high levels of robustness to attack \cite{Heaton:2012a,Fricker:2008b}. In the study on rodent vasculature \cite{Blinder:2010a}, interconnected loops were found to provide robustness to the backbone of surface vessels under edge deletion, and after occlusion to a surface arteriole, the backbone was able to re-rout blood flow and preserve the surrounding neuronal tissue.

In the main text, we assess differences in how the two different types of transport networks balance material costs with network robustness. In particular, we analyze how the size of the largest connected component evolves under the random removal of edges from the network, and take the robustness, $R$, of the network to be the percentage of edges removed in order for the size of the largest connected component to drop to half of its original value. Since we examine random edge removal, there can be some variation in the results when the same procedure is run again on the same network. We thus compute $R$ 20 times for each network in order to generate a representative ensemble average. We should also note that in our analysis, we consider the robustness of the entire vasculature network, rather than just the backbone considered in \cite{Blinder:2010a}. This allows for a cleaner comparison between the vasculature and fungi, and takes into consideration the many nodes that branch directly off the backbone.

The material cost, efficiency, and robustness are all useful quantities to consider in the context of biological networks, but when considered in isolation, it is difficult to understand how they are related to one another and how they compare across similar networks of different size or of completely different type. In order to better understand how the vasculature and fungal systems might be differentially optimized for each of these measures, we used a set of normalized quantities that captured how the wiring cost, efficiency, and robustness of a given network compared to the approximate limiting values for a network of the same size with the same node locations.

In particular, following \cite{Buhl:2004a,Buhl:2006a,Cardillo:2006a}, this is quantified by defining a \textit{relative} cost, efficiency, and robustness. The relative cost is given by
~\\
\begin{equation}
W_{\text{rel}} = \frac{W - W_{\text{MST}}}{W_{\text{GT}} - W_{\text{MST}}},
\label{eq:relW}
\end{equation}
~\\
where $W$, $W_{\text{MST}}$, and $W_{\text{GT}}$ denote the total wiring of the real, \textit{MST}, and \textit{GT}, respectively. In a similar manner, the relative global efficiency, $E_{\text{rel}}$, is given by
~\\
\begin{equation}
E_{\text{rel}} = \frac{E - E_{\text{MST}}}{E_{\text{GT}} - E_{\text{MST}}}.
\label{eq:relE}
\end{equation}
~\\
and the relative robustness $R_{\text{rel}}$ is
~\\
\begin{equation}
\label{eq:relR}
R_{\text{rel}} = \frac{R - R_{\text{MST}}}{R_{\text{GT}} - R_{\text{MST}}}.
\end{equation}
~\\

By definition, the \textit{MST} is minimally wired, but since it contains no shortcuts between nodes, we also expect it to have high average shortest path and low efficiency. It is also clear that the removal of any edge will break the \textit{MST} into disconnected components. On the other hand, the \textit{GT} is a highly expensive network to build, but this increase in material for the same set of nodes should improve the robustness as well as the efficiency of the network, since it allows for shortcuts between pairs of otherwise more distant nodes. Thus, the \textit{MST} is an effective lower bound for material cost, efficiency and robustness, and the \textit{GT} is a good approximation for a planar network that achieves upper bounds with respect to the same measures. The relative measures are normalized between 0 and 1 for the \textit{MST} and \textit{GT}, respectively, and this normalization allows for a direct quantification of where the real distribution networks lie in comparison to the limiting values. Furthermore, the relative quantities can be used to understand and properly contrast the tradeoffs between wiring, efficiency, and resistance to random damage between the vasculature and fungi.


\end{document}